\documentclass[journal,twoside,web]{ieeecolor}
\usepackage{tmi}
\usepackage{cite}
\usepackage{amsmath,amssymb,amsfonts}
\usepackage{algorithmic}
\usepackage{graphicx}
\usepackage{textcomp}
\usepackage{multicol}
\usepackage{multirow}
\usepackage{color}
\usepackage{ulem}
\makeatletter
\let\NAT@parse\undefined
\makeatother
\usepackage[colorlinks=true,linkcolor=blue,citecolor=blue,urlcolor=blue]{hyperref}

\newcommand{\lzt}[1]{#1}
\newcommand{\R}[1]{#1}

\def\BibTeX{{\rm B\kern-.05em{\sc i\kern-.025em b}\kern-.08em
    T\kern-.1667em\lower.7ex\hbox{E}\kern-.125emX}}
\begin{document}

\title{Geometry-Aware Attenuation Learning for Sparse-View CBCT Reconstruction}

\author{{
Zhentao Liu, 
Yu Fang,
Changjian Li, 
Han Wu,
Yuan Liu, 
Dinggang Shen, \IEEEmembership{Fellow, IEEE}, \\
Zhiming Cui
}
\thanks{Zhentao Liu, Yu Fang, Han Wu, Dinggang Shen, and Zhiming Cui are with the School of Biomedical Engineering \& State Key Laboratory of Advanced Medical Materials and Devices, ShanghaiTech Univerisity, Shanghai, 201210, China. Dinggang Shen is also with Shanghai United Imaging Intelligence Co., Ltd., Shanghai, 200230, China, and Shanghai Clinical Research and Trial Center, Shanghai, 201210, China. (e-mail: \{liuzht2022, fangyu, wuhan2022, dgshen, cuizhm\}@shanghaitech.edu.cn). \textit{(Corresponding author: Zhiming Cui.)}}
\thanks{Changjian Li is with the School of Informatics, The University of Edinburgh, Edinburgh, UK (e-mail: chjili2011@gmail.com).}
\thanks{Yuan Liu is with the Department of Computer Science, The University of Hong Kong, Hong Kong, China (e-mail: yuanly@connect.hku.hk).}
\thanks{This work was supported in part by NSFC grants (No. 6230012077). Shanghai Municipal Central Guided Local Science and Technology Development Fund Project no: YDZX20233100001001.}
}

\maketitle

\begin{abstract}

Cone Beam Computed Tomography (CBCT) plays a vital role in clinical imaging. 
Traditional methods typically require hundreds of 2D X-ray projections to reconstruct a high-quality 3D CBCT image, leading to considerable radiation exposure. 
This has led to a growing interest in sparse-view CBCT reconstruction to reduce radiation doses. 
While recent advances, including deep learning and neural rendering algorithms, have made strides in this area, 
\lzt{these methods either produce unsatisfactory results or suffer from time inefficiency of individual optimization.}
\lzt{In this paper, we introduce a novel geometry-aware encoder-decoder framework to solve this problem.}
Our framework starts by encoding multi-view 2D features from various 2D X-ray projections \lzt{with a 2D CNN encoder.} 
\lzt{Leveraging the geometry of CBCT scanning, it then back-projects the multi-view 2D features into the 3D space to formulate a comprehensive volumetric feature map, followed by a 3D CNN decoder to recover 3D CBCT image.}
\lzt{Importantly, our approach respects the geometric relationship between 3D CBCT image and its 2D X-ray projections during feature back projection stage, and enjoys the prior knowledge learned from the data population.}
This ensures its adaptability in dealing with extremely sparse view inputs without individual training, such as scenarios with only 5 or 10 X-ray projections. 
\lzt{Extensive evaluations on two simulated datasets and one real-world dataset demonstrate exceptional reconstruction quality and time efficiency of our method.}

\end{abstract}

\begin{IEEEkeywords}
Sparse-view CBCT reconstruction, geometry awareness, multi-view consistence, prior knowledge.
\end{IEEEkeywords}


\section{Introduction}
\label{sec:introduction}

\begin{figure*}
\centering
    \includegraphics[width=0.98\textwidth]{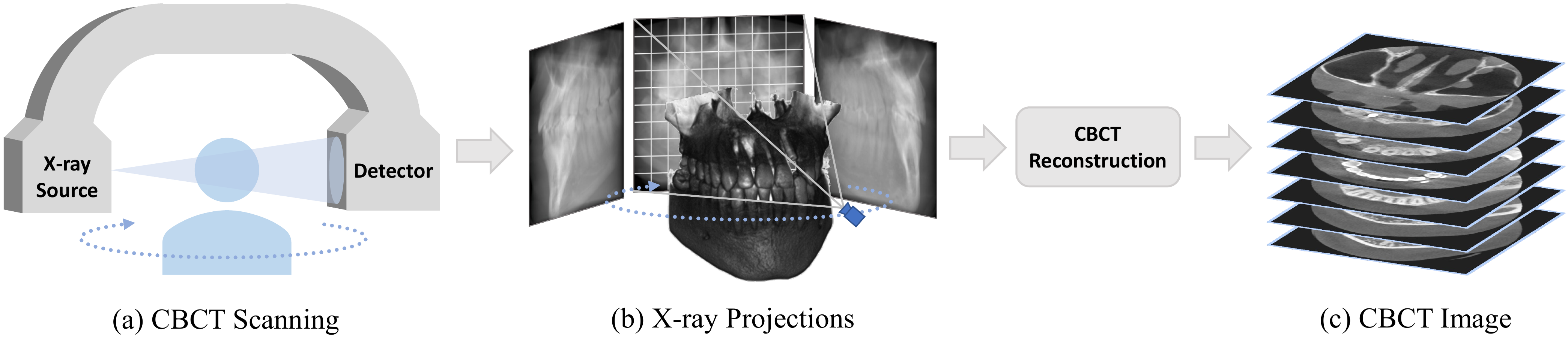}
    \caption{
    \lzt{CBCT scanning and reconstruction. In the CBCT imaging process, CBCT scanning (a) would generate a sequence of 2D X-ray projections (b). These projections are utilized to reconstruct 3D CBCT image (c).}}
    \label{fig0}
    \vspace{-4mm}
\end{figure*}

\IEEEPARstart{C}{one} Beam Computed Tomography (CBCT), a specialized form of CT scanning, is extensively utilized in clinical settings for diagnostic purposes, including dental~\cite{dental_cbct, Funda_cbct}, spinal~\cite{cbct_spine, cbct_spine_1}, and vascular disease diagnosis~\cite{dsa_diagnosis, 4d_dsa}. Compared to the traditional Fan Beam CT (FBCT)~\cite{Funda_cbct}, CBCT offers high-resolution images in a shorter scanning time. A typical CBCT imaging system is illustrated in Fig.~\ref{fig0}. During CBCT scanning, an X-ray source moves uniformly along an arc-shaped trajectory, emitting a cone-shaped beam at each angular step towards the human body, such as the oral cavity. A detector positioned on the patient's opposite side records the 2D projections. Therefore, the CBCT reconstruction essentially transforms into an inverse problem, with the goal of recovering 3D anatomical information from these 2D X-ray projections. However, traditional methods usually require hundreds of projections to produce a high-quality CBCT image [8], raising concerns about radiation exposure. Hence, sparse-view CBCT reconstruction, which reduces the number of projection views to mitigate radiation exposure, has received widespread attention in the research field.

Traditionally, CBCT reconstruction has mainly employed analytical methods like the Filtered Back Projection (FBP) algorithm~\cite{beerslaw,FDK}. While these methods provide rapid reconstruction solutions, they rely heavily on numerous projection views. 
To address these limitations, several iterative optimization-based methods~\cite{SART,TV_recon,TV_recon1,TV_recon2} have been proposed. 
\lzt{However, despite producing improved reconstruction results under sparse input, they are not time efficient and often lack fine details.}
Recently, with the advent of deep learning, \lzt{some} researchers have explored deep learning techniques to learn the mapping between multi-view projections and CBCT \lzt{image in an end-to-end manner} from extensive datasets~\cite{Single-recon, bi-recon, X2CTGAN}.
But these approaches directly concatenate multi-view image information, neglecting the inherent geometric properties of the CBCT system, leading to \lzt{structurally erroneous} reconstruction results. 
Concurrently, in the 3D vision society, the technique of neural rendering (e.g., NeRF~\cite{NeRF}) has garnered significant interest for novel view synthesis and multi-view reconstruction. 
These techniques represent a 3D scene by the neural representation, and employ differentiable rendering for optimization.
\lzt{It is also a typical inverse problem that tries to recover the 3D scene from multi-view image captures, similar to the concept of CBCT reconstruction. 
Our goal is to reconstruct 3D anatomical information from multi-view X-ray projections.}
Building on this, many methods, e.g., \lzt{IntroTomo~\cite{Intratomo}}, NeAT~\cite{NeAT}, NAF~\cite{NAF}, SNAF~\cite{SNAF}, \lzt{SAX-NeRF~\cite{sax_nerf}}, have been developed to incorporate neural rendering into CBCT reconstruction, and achieved impressive results. However, a significant drawback of these methods is that they must be individually optimized for each CBCT scan, making the reconstruction process extremely time-consuming, i.e., using tens of minutes for one patient. More importantly, the quality of reconstructed images diminishes considerably with using the extremely sparse inputs, such as 5 or 10 views.

In this way, sparse-view CBCT reconstruction is a highly ill-posed problem with twofold challenges: (1) How to bridge the dimension gap between multi-view 2D X-ray projections and the CBCT \lzt{image}; (2) How to solve information insufficiency introduced by extremely sparse-view input.  
In this study, we introduce a geometry-aware encoder-decoder framework to solve this task efficiently.
It seamlessly integrates the multi-view consistency of neural rendering \lzt{and} the generalization ability of deep learning, effectively addressing the challenges mentioned above.
Specifically, we first adopt a \lzt{2D convolutional neural network (CNN) encoder} to extract multi-view 2D features from different X-ray projections. 
\lzt{Then, in aligning with the geometry of CBCT scanning, we back-project multi-view 2D features into 3D space,} which properly bridges the dimension gap with multi-view consistency.
Particularly, as different views offer varying degrees \lzt{of information}, an adaptive feature fusion strategy is further introduced to aggregate these multi-view features.
\lzt{Consequently, a 3D volumetric feature is constructed and then decoded into 3D CBCT image with a 3D CNN decoder.}
\lzt{Our framework's inherent geometry awareness ensures accurate information retrieval from multi-view X-ray projections.}
Moreover, by capturing prior knowledge from populations in extensive datasets, our method can generalize well across different patients without individual optimization, even with extremely sparse input views, such as 5 or 10 views. 
\lzt{Experiments on two simulated datasets and one real-world dataset underscore the effectiveness and time efficiency of our method.}


\section{Related Works}
\label{sec:related works}

Sparse-view CBCT reconstruction aims to reduce the number of projections while preserving the quality of the reconstructed CBCT \lzt{image}. Existing methods predominantly fall into three categories: traditional methods, learning-based methods, and neural rendering-based methods, as briefed below.

\subsection{Traditional CBCT Reconstruction}

The Filtered Back Projection (FBP)~\cite{beerslaw} and its 3D cone-beam variant FDK (Feldkamp, Davis, and Kress)~\cite{FDK} stand out as the most prevalent CBCT reconstruction algorithm in the commercial CBCT systems. 
\lzt{FDK back-projects the filtered projection values from different viewpoints into the original 3D space to reconstruct the CBCT image.}
But, to avoid streaks in the CBCT image, it needs hundreds of X-ray projections. This can be harmful because of the high radiation.
To solve this, several iterative methods~\cite{SART, TV_recon, TV_recon1, TV_recon2} have been proposed to leverage the consistency between acquired projections and the reconstructed image. 
For example, SART~\cite{SART} introduces an iterative approach to diminish the disparity between 2D projections and their estimates, updating the \lzt{reconstructed image} at the same time.
Other methods~\cite{TV_recon, TV_recon1, TV_recon2} further incorporate regularization terms to improve reconstruction quality. 
However, while these methods work well in noisy and view-limited scenarios, they usually \lzt{lose image details} and demand more computational resources.

\subsection{Learning-Based CBCT Reconstruction}

\lzt{Deep learning has been increasingly applied to CT/CBCT reconstruction, and could be roughly divided into three classes: data restoration reconstruction~\cite{FBPConvNet, projection-inpaint, DRONE}, model-based iterative reconstruction~\cite{LIR,LIR1,IRON}, and end-to-end reconstruction~\cite{Single-recon,bi-recon,X2CTGAN}.
First, data restoration reconstruction methods mainly include image domain restoration~\cite{FBPConvNet}, sinogram domain restoration~\cite{projection-inpaint}, and dual-domain restoration solutions~\cite{DRONE}.
FBPConvNet~\cite{FBPConvNet} utilizes CNN for image domain restoration to improve upon an initial reconstruction from FBP algorithm.
SinoConvNet~\cite{projection-inpaint} operates in sinogram domain, employing a CNN to inpaint sparse view sinogram to a full view one before performing FBP reconstruction.
DRONE~\cite{DRONE} combines image and sinogram restoration together as a dual-domain approach.
While effective for 2D CT slices, extending these methods to 3D CBCT reconstruction can lead to significant GPU memory overhead. 
Additionally, processing 3D CBCT image slice by slice leads to inter-slice inconsistency.
Second, some model-based iterative methods~\cite{LIR, LIR1} have been developed based on the learned iterative reconstruction technique.
Their basic idea is to unroll an iterative scheme with a deep neural network architecture, explicitly integrating forward projection and back projection into model architecture.
These methods would produce satisfactory results but they may suffer from high-memory footprints and time inefficiency.
\R{IRON~\cite{IRON} is a notable iterative method for limited-angle 2D CT reconstruction.
It adopts a training strategy like image domain restoration~\cite{FBPConvNet} and then deploys the trained sub-network as a regularization prior in the residual image domain for iterative image reconstruction.
Compared to traditional unrolled iterative methods~\cite{LIR, LIR1}, IRON is more memory efficient for avoiding a large number of projection and back projection operations in training.
Its underlying principles could be adapted for sparse-view 3D CBCT reconstruction.}
Third, there are also some end-to-end methods that directly learn the mapping from muti-view projections to CBCT image, as described below.}
PatRecon~\cite{Single-recon} can produce 3D CBCT reconstruction results using just a single X-ray projection.
This framework primarily uses a 2D-to-3D autoencoder architecture. 
A 2D CNN encoder first extracts features from the X-ray projection, which are then reshaped in the feature space to transition into 3D. This is followed by a 3D CNN decoder to produce volumetric image. 
For multiple input views, PatRecon directly concatenates them along the channel dimension, which neglects intrinsic geometric relationships between different projections.
Similarly, Bi-Recon~\cite{bi-recon} and X2CTGAN~\cite{X2CTGAN} operate on \lzt{similar} principle but are limited to handling only two orthogonal projections due to their specific model designs.
They usually produce blurry reconstruction and inaccurate anatomical structures.
These issues arise from their ignorance of the geometric relationship between X-ray projections and CBCT image when combing X-ray projections or their feature representations.

\R{Recently, diffusion models have been widely employed for sparse-view or limited angle CT reconstruction in 2D~\cite{DPS, DOLCE, MOGM, DCDS, TIFA} or 3D format~\cite{diffusionmbir, DDS}.
Their basic idea is to incorporate data consistency term to guide the diffusion model sampling process, aligning the forward projections from the generated image with the input X-ray projections.
MOGM~\cite{MOGM} establishes a multi-channel fusion module for sparse-view CT reconstruction, enhancing the efficiency of data consistency module and providing the diffusion model with more accurate guidance.
DCDS~\cite{DCDS} proposes a collaborative diffusion mechanism for sparse-view CT reconstruction that combines both sinogram and image diffusion model to simultaneously consider dual-domain prior distribution.
TIFA~\cite{TIFA} presents a novel rapid-sampling technique for limited-angle CT reconstruction that incorporates jump-sampling and time-reversion with re-sampling.
This scheme not only accelerates the sampling but also enhances reconstruction outcomes.
TIFA further integrates Diagonal Total Variation (DTV) to mitigate directional artifacts arising from limited-angle input.
DiffusionMBIR~\cite{diffusionmbir} has been utilized for sparse-view 3D CT reconstruction, and DDS~\cite{DDS} offers an accelerated solution with decomposed sampling strategy.
They generate 3D CT volume slice by slice along $z$-axis and use $z$-axis Total Variation (TV) to enforce inter-slice consistency.
However, inter-slice inconsistency still exists.
Besides, their performance has only been validated on parallel beam geometry, and their effectiveness on cone beam geometry requires further verification.}

\subsection{Neural Rendering-Based CBCT Reconstruction}

Neural rendering, including NeRF~\cite{NeRF} and its derivatives~\cite{ACORN, Instant-ngp, Regnerf, pixelnerf, MVSNeRF}
has rapidly progressed in the field of 3D vision. It stands out for its ability to generate new viewpoints and perform multi-view 3D reconstruction, demonstrating significant improvement in maintaining consistency across multiple views. 
\lzt{In X-ray imaging and tomographic reconstruction, many methods have adopted neural rendering for multi-view X-ray synthesizing~\cite{mednerf}, 2D CT reconstruction~\cite{NeRP, SCOPE}, rigid motion correction~\cite{motioncorrection}, metal artifact removal~\cite{polyer}, DSA reconstruction~\cite{TiAVox, VPAL}, and 3D CBCT reconstruction~\cite{Intratomo, NeAT, NAF, SNAF, sax_nerf}.
And in this study, we mainly focus on the methods designed for CBCT reconstruction.
For example, Intratomo~\cite{Intratomo} employs Multilayer Perceptrons (MLPs) with Fourier feature encoding to predict novel sinograms from sparse-view inputs for tomographic reconstruction. 
It then uses geometrical priors to regularize the results.
However, the outputs are often blurry due to the limitations of its encoding module.}
NeAT~\cite{NeAT} uses an octree structure~\cite{ACORN} to represent 3D anatomical structures. 
By combining this with a differentiable rendering algorithm, NeAT achieves excellent reconstruction results. 
Similarly, NAF~\cite{NAF} incorporates a multi-resolution hash table~\cite{Instant-ngp} for \lzt{sparse-view CBCT reconstruction}, showing impressive results even with as few as 50 projection views. 
Its efficient hash data structure helps preserve detailed features, even with limited input views. 
SNAF~\cite{SNAF}, an improvement on NAF, tackles more sparse view conditions (like 20 views) by introducing a new technique for view augmentation. 
It creates additional views between existing ones, which are then used to enhance reconstruction quality.
\lzt{SAX-NeRF~\cite{sax_nerf} achieves plausible results with a line segment-based transformer and masked local-global ray sampling strategy.}
However, all these models require individual training for every subject, taking up to tens of minutes for a single CBCT image reconstruction. 
Moreover, they do not fully overcome the challenges posed by extremely sparse views. 
Their performance tends to drop in quality with extremely limited projections, such as 5 or 10 views.
\R{A notable} deep learning variation of NeRF called PixelNeRF~\cite{pixelnerf} was introduced, aiming to tackle the sparse-view challenge by using scene knowledge learned from large datasets. 
Another concurrent approach, DIF-Net~\cite{DIF-Net}, applies this concept to medical imaging and shows promising results on a knee CBCT dataset. 
\lzt{However, it uses MLPs for point-wise decoding, which overlooks the interactions between neighboring points of CBCT image.
And the point-wise 3D supervision makes it difficult to capture the global structure information of CBCT image.
As a result, it exhibits streaky artifacts when dealing with extremely sparse input, like 5 views.}

Our approach combines the advantages of deep learning and neural rendering, providing both generalization \lzt{capability} and multi-view consistency. 
Remarkably, our framework is capable of delivering reliable reconstruction quality \lzt{using 20 input views in a second}.
This highlights superior performance and \lzt{time} efficiency of our method.


\section{Method}

In this section, we first introduce data preparation in Sec.~\ref{sec:dataacq}. \lzt{Following that}, we elaborate our methodology in detail in Sec.~\ref{sec:field}.

\subsection{Data Preparation}
\label{sec:dataacq}

As shown in Fig.~\ref{fig0}, the X-ray source moves around the patient along a set arc-shaped path. 
Opposite this source, there is a 2D detector that captures X-ray projections of the body, like the oral cavity. 
\lzt{In real-world collection, initial X-ray photons emitted from the source are attenuated differently by various tissues in the body.
The detector recorded X-ray photon count would be converted into X-ray projection value after flat-field and dark-field corrections~\cite{walnuts}.
The projection value represents the line integral of attenuation along the X-ray path.
For a detailed derivation of X-ray attenuation process, please refer to Sec.~\ref{sec:dataacq-realworld}.
Throughout this paper, the term \textbf{X-ray projection} specifically refers to the attenuation ray integral for clarity.
The attenuation value is shown as voxel intensity value we see in CBCT image.
Therefore, our objective is to reconstruct the 3D CBCT image, using a limited set of X-ray projections.
In our experiments, we adopt two simulated datasets of dental and spine~\cite{ctspine1k} and one real-world dataset of walnut~\cite{walnuts} to verify our method performance. 
In the following, we introduce how we prepare our X-ray and CBCT paired data.
The detailed specifics of dataset will be provided in Sec.~\ref{sec:dataset}.}

\subsubsection{Simulated X-ray and CBCT Pairs Collection}
\label{sec:dataacq-simulate}

\begin{figure}
\centering
    \includegraphics[width=0.48\textwidth]{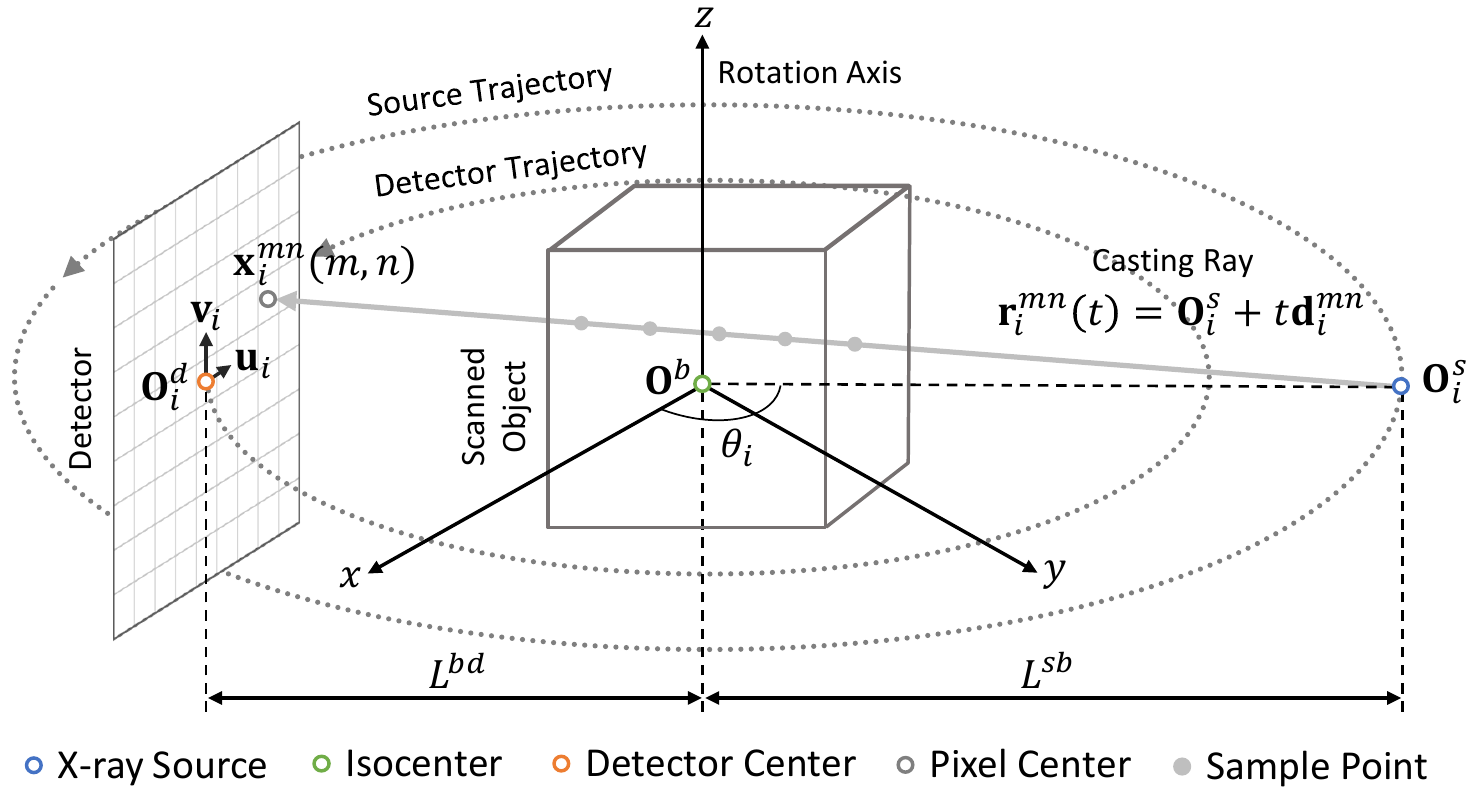}
    \caption{
    \lzt{Geometric configuration of CBCT scanning and X-ray projection simulation.}
    } 
    \label{fig8}
    \vspace{-4mm}
\end{figure}

\lzt{For simulated datasets, we use the Digitally Reconstructed Radiography (DRR) technique to simulate multiple X-ray projections from a given CBCT image.}

\lzt{A typical geometric configuration of CBCT scanning is illustrated in Fig.~\ref{fig8}.
The isocenter $\mathbf{O}^{b}\in\mathbb{R}^3$ is the center point of the scanned object.
We establish the world coordinate system with $\mathbf{O}^{b}$ as the origin, i.e., $\mathbf{O}^{b}=[0,0,0]^{T}$, aligning the axes with the bounding box of the scanned object.
Given a volumetric CBCT image $\mathbf{V}_{gt}\in\mathbb{R}^{1 \times W \times H \times D}$ as the scanned object with a spatial resolution of $W \times H \times D$ and a voxel spacing of $v_x \times v_y \times v_z$, we define the volumetric coordinate matrix $\mathcal{V}\in\mathbb{R}^{3\times W \times H \times D}$ that bounds the CBCT image in world space.
Each element from $\mathcal{V}$ represents the center coordinate of voxel grid from CBCT image.
The X-ray source and the detector plane rotate around $z$-axis simultaneously within the range of $[0^\circ, 360^\circ)$, completing a full scan.
Using a consistent angular step $\Delta\theta=18^\circ$, this setup creates 20 X-ray projections.
For other configurations, like 10 or 5 views, we use larger angular steps of $36^\circ$ and $72^\circ$, respectively.
The distance between X-ray source and object is denoted as $L^{sb}$ and the distance between object and detector is denoted as $L^{bd}$.
In the following, we introduce the simulation of the $i$-th X-ray projection, where $i \in \{1,2,\ldots,N\}$ is the view index and $N$ is the total number of views.}

\lzt{For the $i$-th viewpoint, our objective is to simulate X-ray projection $\mathbf{P}_{i}\in\mathbb{R}^{1\times w\times h}$ with a spatial resolution of $w\times h$ and a pixel spacing of $p_u \times p_v$ from the given CBCT image.
The rotation angle is defined as $\theta_i=\Delta\theta (i-1)$. Then, the position of the X-ray source $\mathbf{O}^{s}_{i}\in\mathbb{R}^3$ and the position of the detector center $\mathbf{O}^{d}_{i}\in\mathbb{R}^3$ could be given by:
\begin{equation}
    \mathbf{O}^{s}_{i} = \begin{bmatrix}
    L^{sb}\cos{\theta_i} \\
    L^{sb}\sin{\theta_i} \\
    0
    \end{bmatrix},~ 
    \mathbf{O}^{d}_{i} = \begin{bmatrix}
        -L^{bd}\cos{\theta_i} \\
        -L^{bd}\sin{\theta_i} \\
        0
    \end{bmatrix}
    \label{eq:xraysource, detectorcenter}
\end{equation}
Eq.~\ref{eq:xraysource, detectorcenter} is derived under the assumption that $\mathbf{O}^{s}_{i}$, $\mathbf{O}^{d}_{i}$, and $\mathbf{O}^{b}$ are colinear in the $xy$-plane, with the $z$-axis values of $\mathbf{O}^{s}_{i}$ and $\mathbf{O}^{d}_{i}$ both being zeros~\cite{comprehensivephysics}.
The detector 3D basis vector between adjacent pixels in horizontal direction $\mathbf{u}_{i}\in\mathbb{R}^3$ and the basis vector in vertical direction $\mathbf{v}_{i}\in\mathbb{R}^3$ could be given by:
\begin{equation}
    \mathbf{u}_{i} = \begin{bmatrix}
        -p_u\sin{\theta_i} \\
        p_u\cos{\theta_i} \\
        0
    \end{bmatrix},~
    \mathbf{v}_{i} = \begin{bmatrix}
        0 \\
        0 \\
        p_v
    \end{bmatrix}
    \label{eq:uspace, vspace}
\end{equation}
Given a pixel $(m,n)$ on detector plane, where $m\in\{0,1,\ldots,w-1\}, n\in\{0,1,\ldots,h-1\}$, we aim to derive its pixel center coordinate $\mathbf{x}^{mn}_i\in\mathbb{R}^3$ in world space.
And we first derive the pixel center coordinate $\mathbf{x}^{00}_i\in\mathbb{R}^3$ of pixel $(0,0)$ as follows, suitable for both odd and even image resolution:
\begin{align}
    \mathbf{x}^{00}_i = \mathbf{O}^{d}_{i} 
    &- \left( \left\lfloor \frac{w}{2} \right\rfloor + \left\lfloor \frac{w+1}{2} \right\rfloor - \frac{w+1}{2} \right)\mathbf{u}_{i} \notag \\
    &- \left( \left\lfloor \frac{h}{2} \right\rfloor + \left\lfloor \frac{h+1}{2} \right\rfloor - \frac{h+1}{2} \right)\mathbf{v}_{i}
\end{align}
where $\lfloor\cdot\rfloor$ denotes the flooring operation. 
Then $\mathbf{x}^{mn}_{i}$ could be given by:
\begin{align}
    \mathbf{x}^{mn}_i = \mathbf{x}^{00}_i + m\mathbf{u}_{i} + n\mathbf{v}_{i}
\end{align}
Consider an incident X-ray path, $\mathbf{r}^{mn}_{i}(t)=\mathbf{O}^s_i+t\mathbf{d}^{mn}_{i}\in\mathbb{R}^3$, radiating from the X-ray source $\mathbf{O}^s_i$ towards the $\mathbf{x}^{mn}_i$ on the detector, and $t\in\mathbb{R}^{+}_{0}$ is the scaling parameter.
The direction vector $\mathbf{d}^{mn}_{i}\in\mathbb{R}^3$ could be formulated as:
\begin{equation}
    \mathbf{d}^{mn}_{i} = \mathbf{x}^{mn}_i - \mathbf{O}^s_i
\end{equation}
This ray intersects the bounding box of the CBCT volume at entry point $\mathbf{r}^{mn}_{i}(t_n)$ and exit point $\mathbf{r}^{mn}_{i}(t_f)$ according to Ray-AABB algorithm~\cite{rayaabb}.
Here, $t_{n}$ and $t_{f}$ denote the near and far bounds of the volume, respectively.
We then simulate the ray integral of the attenuation value along the ray:
\begin{equation}
\hat{P}(\mathbf{r}^{mn}_{i}\lvert\mathbf{V}_{gt}) = \int_{t_n}^{t_f}\hat{\mu}(\mathbf{r}^{mn}_{i}(t)\lvert\mathbf{V}_{gt})\mathrm{d}t
\label{continous simulated ray integral}
\end{equation}
where $\hat{P}(\mathbf{r}^{mn}_{i}\lvert\mathbf{V}_{gt})\in\mathbb{R}^{+}_{0}$ is the simulated projection value. 
$\hat{\mu}:\left(\mathbb{R}^{3} \lvert \mathbb{R}^{1\times W \times H \times D} \right)\rightarrow\mathbb{R}^{+}_{0}$ is a function that describes the attenuation distribution of the given CBCT image, and $\hat{\mu}(\mathbf{r}^{mn}_{i}(t)\lvert\mathbf{V}_{gt})$ is the attenuation value for point $\mathbf{r}^{mn}_{i}(t)$, defined as follows:
\begin{equation}
\hat{\mu}(\mathbf{r}^{mn}_{i}(t)\lvert\mathbf{V}_{gt})=\mathrm{Interp}_k\left(\mathbf{V}_{gt}, \mathbf{r}^{mn}_{i}(t)\right), k=3
\end{equation}
where $\mathrm{Interp}_k:\left( \mathbb{R}^{C_1\times D_1 \times D_2 \times \cdots \times D_k}, \mathbb{R}^{k} \right)\rightarrow\mathbb{R}^{C_1}$ is $k$-linear interpolation.
Here $C_1=1$ represents the attenuation value, and we use trilinear interpolation in the above equation.
For computational feasibility, we discretize Eq.~\ref{continous simulated ray integral} by uniformly sampling points between the near and far bounds as shown in Fig.~\ref{fig8}, deriving the following equation:
\begin{equation}
\hat{P}(\mathbf{r}^{mn}_{i}\lvert\mathbf{V}_{gt}) = \sum_j\hat{\mu}(\mathbf{r}^{mn}_{i}(t_j)\lvert\mathbf{V}_{gt})\delta_j
\label{discrete simulated ray integral}
\end{equation}
where $\mathbf{r}^{mn}_{i}(t_j)$ represents the $j$-th sampling point, and $\delta_j=0.5\min(v_x,v_y,v_z)$ is the distance between adjacent points.}

\lzt{By combining the projection values of all pixels on the detector, we can finally simulate the X-ray projection $\mathbf{P}_i$ for the $i$-th viewpoint.
Similarly, we can synthesize X-ray projections from all viewpoints $\{\mathbf{P}_i\}_{i=1}^{N}$ for the given CBCT volume $\mathbf{V}_{gt}$. 
In this way, we obtain an X-ray and CBCT paired data.
For each of dental and spine dataset, we collect 130 CBCT/CT volumes, and simulate the X-ray projections as mentioned above.}

\subsubsection{Real-World X-ray and CBCT Pairs Collection}
\label{sec:dataacq-realworld}

\lzt{However, the noise-free X-ray projections from our simulated dataset are always too ideal.
In real-world collection, X-ray projections inevitably include Possion noise, and electronic noise introduced by the imaging system.
To verify our robustness, we further collect another real-world captured CBCT walnut dataset~\cite{walnuts}.
In the following, we provide a detailed formulation of X-ray attenuation process in real-world collection, and the details of the walnut dataset.}

\lzt{Similarly, imaging an X-ray path $\mathbf{r}^{mn}_{i}(t)$ as we defined in Sec.~\ref{sec:dataacq-simulate}, the X-ray attenuation process along this path can be expressed according to Beer's Law~\cite{beerslaw}:
\begin{align}
    I(\mathbf{r}^{mn}_{i}) = & \left(I_0(\mathbf{r}^{mn}_{i}) - I_1(\mathbf{r}^{mn}_{i})\right) \exp{\left(-\int \mu(\mathbf{r}^{mn}_{i}(t)) \, \mathrm{d}t \right)} \notag \\
    & + I_1(\mathbf{r}^{mn}_{i})
\end{align}
here $I(\mathbf{r}^{mn}_{i})\in\mathbb{Z}^{+}_{0}$ denotes the X-ray photon count recorded by the detector, $I_0(\mathbf{r}^{mn}_{i})\in\mathbb{Z}^{+}_{0}$ denotes the X-ray source emitted X-ray photon count (flat-field image), and $I_1(\mathbf{r}^{mn}_{i})\in\mathbb{Z}^{+}_{0}$ denotes the detector offset count (dark-field image). 
All these quantities are pixel-dependent.
$\mu:\mathbb{R}^{3}\rightarrow\mathbb{R}^{+}_{0}$ is a function that describes the attenuation distribution of the scanned scene, and $\mu(\mathbf{r}^{mn}_{i}(t))$ is attenuation value for point $\mathbf{r}^{mn}_{i}(t)$.
By conducting flat-field and dark-field corrections~\cite{walnuts}, we would get the X-ray projection value as follows:
\begin{equation}
    P(\mathbf{r}^{mn}_{i}) = -\ln \left( \frac{I(\mathbf{r}^{mn}_{i})-I_1(\mathbf{r}^{mn}_{i})}{I_0(\mathbf{r}^{mn}_{i})-I_1(\mathbf{r}^{mn}_{i})}\right) = \int \mu(\mathbf{r}^{mn}_{i}(t)) \mathrm{d}t
    \label{flat-dark-corretion}
\end{equation}}

\lzt{In this dataset, each walnut was scanned three times with the X-ray source set at three different heights: high, middle, and low position.
Each scan comprises 1201 views with a uniform increment of $0.3^\circ$ within a range of $[0^\circ,360^\circ]$, where the first and the last views are captured at the same location.
For each scan, flat-field and dark-field images were also recorded to pre-process the raw photon count data, as defined in Eq.~\ref{flat-dark-corretion}.
The processed X-ray projections from three scans were combined together to obtain a reference ground-truth CBCT reconstruction that is free from cone angle artifacts.
In our experiments, we utilize $N$ uniformly spaced X-ray projections $\{\mathbf{P}_{i}\}_{i=1}^{N}$ selected from middle scan and the reference CBCT volume $\mathbf{V}_{gt}$ as paired data.
We use $N=20, 10, 5$, resulting in angular increments of $18^\circ$, $36^\circ$, and $72^\circ$, respectively. 
Note that in this dataset, CBCT scanning geometry parameters including $\mathbf{O}^{s}_{i}$, $\mathbf{O}^d_{i}$, $\mathbf{u}_{i}$, and $\mathbf{v}_{i}$ are not calculated by Eqs.~\ref{eq:xraysource, detectorcenter}--\ref{eq:uspace, vspace}, but are recorded in geometry description files provided by~\cite{walnuts}.}

\subsection{Geometry-Aware Attenuation Learning}
\label{sec:field}

\begin{figure*}
\centering
    \includegraphics[width=0.95\textwidth]{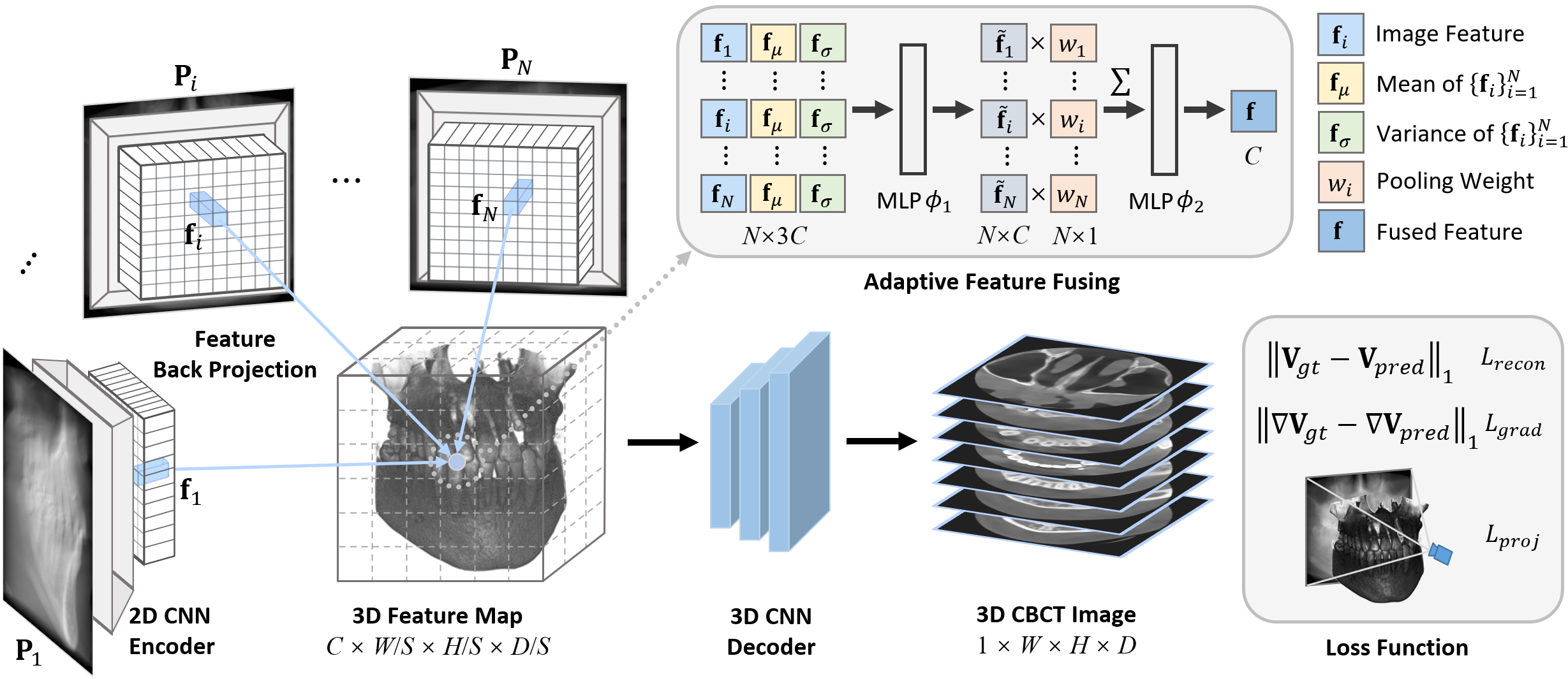}
    \caption{
    \lzt{Overview of our proposed method. A 2D CNN encoder first extracts feature representations from multi-view X-ray projections. 
    Then, we build a 3D feature map by feature back projection and adaptive feature fusing. 
    Finally, this 3D feature map is fed into a 3D CNN decoder to produce the final CBCT image.}
    } 
    \label{fig1}
    \vspace{-4mm}
\end{figure*}

Fig.~\ref{fig1} shows the design of our proposed method.
\lzt{Given a set of sparse-view X-ray projections $\{\mathbf{P}_{i}\}^N_{i=1}$, our objective is to reconstruct the CBCT image $\mathbf{V}_{pred}\in\mathbb{R}^{1 \times W \times H \times D}$ to be as close as possible to the ground truth one $\mathbf{V}_{gt}$.}
We first extract their features using a shared 2D CNN encoder. 
\lzt{Then, we build the 3D feature map by combining feature back projection module with an adaptive feature fusion process.}
After this, the 3D feature map is processed by a \lzt{3D CNN decoder to recover the target CBCT image.}
In particular, one key step of our method is the \lzt{feature back projection}, which effectively bridges the dimensional gap between 2D X-ray projections and 3D CBCT \lzt{image}.
\lzt{It is crucial for achieving anatomically precise reconstructions.}
\lzt{Secondly}, the prior knowledge learned from datasets by deep learning, equips our model with the capability to adapt to new patients without \lzt{individual optimization.} 
\lzt{It is particularly beneficial in handling sparse-view inputs with limited information,  especially when dealing with 5 or 10 views.} 
\lzt{These two factors are the primary reasons why our model could produce satisfactory results with sparse input.}
\lzt{Furthermore, our volume-wise CNN decoder acts as a learnable filter to reduce noise and extract more robust feature representations. 
It helps us better capture the global structure information of the target CBCT image, mitigating streaky artifacts.}
We will delve into the details of our method in the following.

\subsubsection{2D Feature Extraction} 
\label{feature extraction}
Given a set of X-ray projections \lzt{$\{\mathbf{P}_{i}\}^N_{i=1}$ that we defined in Sec.~\ref{sec:dataacq}}, we employ a shared 2D CNN encoder to extract a 2D feature map for each projection. This process allows us to capture view-specific feature representations, expressed as \lzt{$\{\mathbf{F}_i\}^N_{i=1}\subset\mathbb{R}^{C \times w \times h}$}, where $C$ is the feature channel size.

\subsubsection{Feature Back Projection} 
\label{feature_bp}

\begin{figure}
\centering
    \includegraphics[width=0.48\textwidth]{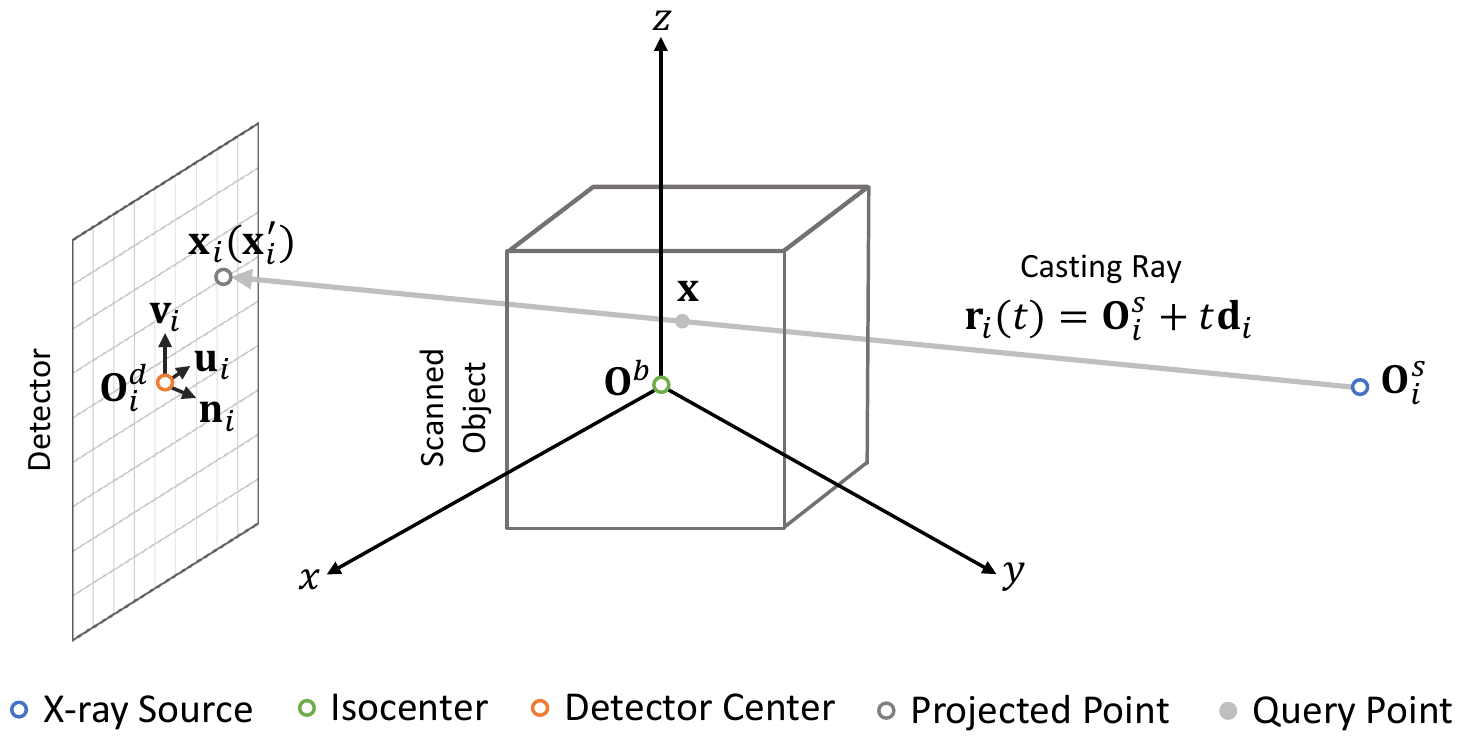}
    \caption{
    \lzt{Coordinate transformation of query point for feature back projection.}
    } 
    \label{forwardprojection}
    \vspace{-4mm}
\end{figure}

\lzt{The key of our framework is feature back projection, crafted to align with the geometry of CBCT imaging system.
The idea is to retrieve the view-specific pixel-aligned feature representation $\mathbf{f}_i\in\mathbb{R}^C$ from $\mathbf{F}_{i}$ given any 3D query point $\mathbf{x}\in\mathbb{R}^3$ in world space, where $\mathbf{x}$ is sampled from volumetric coordinate matrix $\mathcal{V}$, i.e., $\mathbf{x}\in\mathcal{V}$.
We first need to project $\mathbf{x}$ onto detector plane as view-specific projected point $\mathbf{x}_i\in\mathbb{R}^3$ and then transfer it into pixel $\mathbf{x}_i'\in\mathbb{R}^2$ for bilinear interpolation from $\mathbf{F}_{i}$.
This process leverages the CBCT scanning geometry parameters, including the source position $\mathbf{O}^{s}_{i}$, the detector center $\mathbf{O}^d_{i}$, and the detector plane basis vectors $\mathbf{u}_{i}$ and $\mathbf{v}_{i}$ that we mentioned in Sec.~\ref{sec:dataacq}.}

\lzt{In the following, we provide mathematically detailed process of feature back projection. 
A visual illustration is given in Fig.~\ref{forwardprojection}.
If an X-ray path $\mathbf{r}_i(t)=\mathbf{O}^s_{i}+t\mathbf{d}_i\in\mathbb{R}^{3}$ radiating from X-ray source position $\mathbf{O}^s_{i}$ and passing through $\mathbf{x}$, it will intersect the detector plane with projected point $\mathbf{x}_{i}$ for the $i$-th view.
The ray direction $\mathbf{d}_i\in\mathbb{R}^{3}$ could be denoted as:
\begin{equation}
    \mathbf{d}_i = \mathbf{x} - \mathbf{O}^s_i
\end{equation}
The normal vector $\mathbf{n}_{i}$ of detector plane could be denoted as:
\begin{equation}
    \mathbf{n}_{i} = \mathbf{u}_{i} \times \mathbf{v}_{i}
\end{equation}
$\mathbf{x}_{i}$ would satisfy the following plane function as it belongs to the detector plane:
\begin{equation}
    \mathbf{n}_{i} \cdot (\mathbf{x}_{i}-\mathbf{O}^d_{i}) = 0
    \label{eq:plane function}
\end{equation}
and $\mathbf{x}_{i}$ also satisfy the ray function:
\begin{equation}
    \mathbf{x}_{i} = \mathbf{O}^{s}_{i} + t_{\mathbf{x}_i} \mathbf{d}_i
    \label{eq:ray function}
\end{equation}
Substitute Eq.~\ref{eq:ray function} into Eq.~\ref{eq:plane function}, we could get:
\begin{equation}
    \mathbf{n}_{i} \cdot (\mathbf{O}^{s}_{i} - \mathbf{O}^{d}_{i}) + t_{\mathbf{x}_i}(\mathbf{n}_{i}\cdot\mathbf{d}_i) = 0
\end{equation}
In this way, we could get the scaling parameter $t_{\mathbf{x}_i}$ for point $\mathbf{x}_i$ as:
\begin{equation}
    t_{\mathbf{x}_i} = -\frac{\mathbf{n}_{i} \cdot (\mathbf{O}^{s}_{i} - \mathbf{O}^{d}_{i})}{\mathbf{n}_{i}\cdot\mathbf{d}_i}
\end{equation}
Then $\mathbf{x}_{i}$ could be represented using $\mathbf{x}$, $\mathbf{O}^{s}_{i}$, $\mathbf{O}^{d}_{i}$, $\mathbf{u}_{i}$, and $\mathbf{v}_{i}$:
\begin{equation}
    \mathbf{x}_{i} = \mathbf{O}^{s}_{i} 
    - \frac{(\mathbf{u}_{i} \times \mathbf{v}_{i}) \cdot (\mathbf{O}^{s}_{i} - \mathbf{O}^{d}_{i})}
           {(\mathbf{u}_{i} \times \mathbf{v}_{i}) \cdot \left(\mathbf{x} - \mathbf{O}^{s}_{i}\right)}
      \left(\mathbf{x} - \mathbf{O}^{s}_{i}\right)
\end{equation}}
\lzt{Following that, we transform $\mathbf{x}_{i}$ into pixel $\mathbf{x}_{i}'$:
\begin{equation}
    \mathbf{x}_{i}' = 
    \begin{bmatrix}
        (\mathbf{x}_{i}-\mathbf{x}^{00}_{i})\cdot\mathbf{u}_i/\|\mathbf{u}_i\|^2_{2}\\
        (\mathbf{x}_{i}-\mathbf{x}^{00}_{i})\cdot\mathbf{v}_i/\|\mathbf{v}_i\|^2_{2} 
    \end{bmatrix}
\end{equation}
Then, we could get view-specific pixel-aligned features $\mathbf{f}_i$ from $\mathbf{F}_{i}$ by bilinear interpolation:
\begin{equation}
\mathbf{f}_i=\mathrm{Interp}_k\left(\mathbf{F}_{i}, \mathbf{x}_{i}'\right), k=2
\end{equation}
Similarly, we obtain multi-view feature vectors $\{\mathbf{f}_i\}_{i=1}^N$ for a query point $\mathbf{x}$ from different 2D feature maps $\{\mathbf{F}_i\}_{i=1}^N$.}

\lzt{This design bridges the dimension gap between 2D X-ray projections and 3D CBCT image and facilitates accurate information retrieval from multi-view X-ray projections for 3D spatial queries, which is crucial for achieving anatomically precise reconstructions.
It distinguishes our method from previous end-to-end learning-based methods\cite{bi-recon,Single-recon,X2CTGAN} that brutally concatenate information of different views.}

\subsubsection{Adaptive Feature Fusing} 
\label{adaptive_fusing}
After gathering the multi-view feature vectors $\{\mathbf{f}_i\}_{i=1}^N$ for a query point $\mathbf{x}$ through feature back projection, our goal is to merge these feature vectors into a point-wise feature vector $\mathbf{f}\in\mathbb{R}^C$.
\lzt{Different view X-ray projections present different information due to the variations in X-ray attenuation process introduced by varying viewpoints.
To account for this property during the feature merging process, we draw inspiration from~\cite{IBRNet} and employ an adaptive feature fusion mechanism to effectively integrate these feature vectors.}

For the multi-view feature vectors $\{\mathbf{f}_i\}_{i=1}^N$ associated with the query point $\mathbf{x}$, we start by calculating an element-wise average vector $\mathbf{f}_\mu \in \mathbb{R}^C$ and a variance vector $\mathbf{f}_\sigma \in \mathbb{R}^C$ to capture global \lzt{information across $N$ views:
\begin{equation}
    \mathbf{f}_\mu = \frac{1}{N} \sum_{i=1}^N \mathbf{f}_i,~\mathbf{f}_\sigma = \frac{1}{N} \sum_{i=1}^N (\mathbf{f}_i - \mathbf{f}_\mu)^2
\end{equation}
Each $\mathbf{f}_i$ represents local information retrieved from the $i$-th X-ray projection, while $\mathbf{f}_\mu$ and $\mathbf{f}_\sigma$ capture the common characteristics and variabilities across different viewpoints.
These local features are concatenated with $\mathbf{f}_\mu$ and $\mathbf{f}_\sigma$, and then fed into the first MLP $\phi_1:\mathbb{R}^{3C} \rightarrow \mathbb{R}^{C+1}$ to merge both local and global information:
\begin{equation}
    \tilde{\mathbf{f}}_i, \tilde{w}_i = \phi_1\left(\mathbf{f}_i \oplus \mathbf{f}_\mu \oplus \mathbf{f}_\sigma\right),~i=1,2,\ldots,N
    \label{firstmlpfusion}
\end{equation}
here $\oplus$ denotes concatenation operation. 
Note that in this context, the local or global information represents the single-view or cross-view information for this specific query point $\mathbf{x}$.
In this way, local feature $\mathbf{f}_i$ could interact with other views information through global features $\mathbf{f}_\mu, \mathbf{f}_\sigma$ with the help of $\phi_1$.
This process produces a refined feature vector $\tilde{\mathbf{f}}_i\in \mathbb{R}^C$ and an unnormalized weight $\tilde{w}_{i}\in\mathbb{R}$ for each view.
And $\tilde{w}_{i}$ further undergoes Softmax normalization to obtain a normalized weight ${w}_{i}\in[0,1]$ as follows:
\begin{equation}
    w_i = \frac{\exp(\tilde{w}_i)}{\sum_{l=1}^N \exp(\tilde{w}_l)},~ i=1,2,\ldots,N
\end{equation}
Finally, $\{\tilde{\mathbf{f}}_i\}_{i=1}^{N}$ are summed together using their respective weights $\{{w}_{i}\}^{N}_{i=1}$ and passed through the second MLP $\phi_2:\mathbb{R}^{C} \rightarrow \mathbb{R}^{C}$, resulting in the well-integrated feature vector $\mathbf{f}$:
\begin{equation}
\mathbf{f} = \phi_2\left( \sum_{i=1}^N w_i \tilde{\mathbf{f}}_i \right)
\end{equation}
The weight $w_i$ reflects the relative importance of the $i$-th view, and both $\phi_1$ and $\phi_2$ contain one neuron layer with GELU output activation.}

This fusion technique excels at integrating features from various views, enhancing the ability to capture fine details in regions of CBCT images with typically low contrast.

\subsubsection{Attenuation Decoding}
\label{att_decode}
After obtaining the feature vector $\mathbf{f}$ for each query point \lzt{$\mathbf{x}\in\mathcal{V}$}, we compose these vectors to form the \lzt{volumetric feature map}.
To mitigate GPU memory overhead and speed up computations, we employ a sparse sampling technique \lzt{on $\mathcal{V}$ with a downsampling rate $S\in\{2^a\lvert a\in\mathbb{Z}^{+}\}$}.
This approach enables us to generate a lower-resolution \lzt{3D feature map $\mathcal{F}\in\mathbb{R}^{C\times \frac{W}{S} \times \frac{H}{S} \times \frac{D}{S}}$.
This assembled 3D feature map serves as the feature representation for the target CBCT image.
We then feed it into our 3D CNN decoder to recover the CBCT image $\mathbf{V}_{pred}$, decoding the feature representations as attenuation values.}

\lzt{Our CNN-based decoder considers interactions among neighboring query points.
This approach effectively acts as a learnable filter to mitigate noise and extracts more robust feature representations.
And the volume-wise 3D supervision will help our model capture the global structural information of the target CBCT image.
Consequently, our reconstructions exhibit high quality with less noises and streaky artifacts.}

\subsubsection{Model Optimization}

To effectively train our framework, we incorporate several loss terms to guide and supervise the training process.
\lzt{First, we define the reconstruction loss $L_{recon}$ to enforce voxel-wise similarity between the ground-truth $\mathbf{V}_{gt}$ and the prediction $\mathbf{V}_{pred}$:
\begin{equation}
    L_{recon} = \| \mathbf{V}_{gt} - \mathbf{V}_{pred} \|_1
\end{equation}}To capture finer details, we introduce a gradient loss $L_{grad}$:
\lzt{\begin{equation}
    L_{grad} = \| \nabla\mathbf{V}_{gt} - \nabla\mathbf{V}_{pred} \|_1
\end{equation}
here $\nabla$ denotes the first derivative.
To make the model output align with the input X-ray projections, we further incorporate projection loss $L_{proj}$.
For simulated dataset, $L_{proj}$ is defined as follows:
\begin{equation}
    L_{proj} = \sum_{\mathbf{r}^{mn}_{i}\in\mathbf{B}} \Vert \hat{P}(\mathbf{r}^{mn}_{i}\lvert\mathbf{V}_{gt}) - \hat{P}(\mathbf{r}^{mn}_{i}\lvert\mathbf{V}_{pred}) \Vert_1,
\end{equation}
where $\hat{P}(\mathbf{r}^{mn}_{i}\lvert\mathbf{V}_{pred})$ could be easily derived from Eq.~\ref{discrete simulated ray integral}, and $\mathbf{B}\subset\mathbb{R}^3$ is the random sampled ray batch set from input views.
As for real-world dataset, $L_{proj}$ is defined as follows:
\begin{equation}
    L_{proj} = \sum_{\mathbf{r}^{mn}_{i}\in\mathbf{B}} \Vert P(\mathbf{r}^{mn}_{i}) - \hat{P}(\mathbf{r}^{mn}_{i}\lvert\mathbf{V}_{pred}) \Vert_1,
\end{equation}}Therefore, our final objective function is defined as:
\begin{equation}
    L = L_{recon} + \lambda_{grad}L_{grad} + \lambda_{proj}L_{proj},
\end{equation}
where $\lambda_{grad}$ and $\lambda_{proj}$ determine the relative importance of the gradient and projection loss terms, respectively.

\section{Experiment}

\subsection{Experimental Settings}
\label{sec:exp-set}
\subsubsection{Dataset} 
\label{sec:dataset}

To evaluate the effectiveness of our framework, we conducted experiments using both \lzt{simulated (dental and spine~\cite{ctspine1k}) and real-world dataset (walnut~\cite{walnuts})}. Below are the specifics of these datasets.

For the dental dataset, we collected 130 dental CBCT \lzt{images} from various patients. 
Each \lzt{image} is characterized by a resolution of $256\times256\times256$ and a voxel size of $\mathrm{0.3133mm\times0.3133mm\times0.3133mm}$.
Of these \lzt{images}, 100 \lzt{images} are used for training, 10 for validation, and the remaining 20 for testing.
In our experiments, we test the reconstruction of CBCT images under three different input scenarios, i.e., 5 views, 10 views, and 20 views, respectively.
The corresponding X-ray projections, following the procedure described in Sec.~\ref{sec:dataacq-simulate}, have a resolution of $256\times256$ and a pixel size of $\mathrm{0.4386mm\times0.4386mm}$.
\lzt{The source-to-object distance $L^{sb}=\mathrm{500mm}$, and the object-to-detector distance $L^{bd}=\mathrm{200mm}$.}

Our second dataset includes 130 spinal CT \lzt{images} sourced from CTSpine1K\cite{ctspine1k}.
It is worth noting that spinal \lzt{images} typically encompass a variety of organs and soft tissues, often with limited contrast in intensity. 
This aspect poses significant challenges for sparse-view reconstruction, a task notably more challenging than dental \lzt{images}. 
Hence, including spinal \lzt{images} in our study enables us to verify robustness and versatility of our framework across different anatomical areas.
To ensure uniformity across \lzt{cases}, we resampled, cropped, and padded each spinal CT \lzt{image} to a resolution of $256\times256\times256$ and a voxel size of $\mathrm{2mm\times2mm\times2mm}$.
We split the spinal \lzt{images} into three groups: 100 \lzt{images} for training, 10 for validation, and 20 for testing.
We also replicated the same input view configurations (namely, 5, 10, and 20 projections) for each \lzt{case}, as detailed in Sec.~\ref{sec:dataacq-simulate}. 
Each projection has a resolution of $256\times256$ with a pixel size of $\mathrm{3mm\times3mm}$.
\lzt{The source-to-object distance $L^{sb}=\mathrm{1000mm}$, and the object-to-detector distance $L^{bd}=\mathrm{500mm}$.}

\lzt{To further verify our robustness in real-world setting, we collect the third dataset includes 42 real-world walnut CBCT scanning collections from~\cite{walnuts}.
We split this dataset into 32 cases for training, 5 cases for validation, and 5 cases for testing.
The original X-ray projection has a resolution of $768\times972$ and a pixel size of $0.1496\mathrm{mm}\times0.1496\mathrm{mm}$.
We downsample the projection to a resolution of $256\times324$ and a pixel size of $0.4488\mathrm{mm}\times0.4488\mathrm{mm}$.
For each case, we use the downsampled X-ray projections from three positioned scans to reconstruct the reference CBCT image based on the iterative algorithm from~\cite{walnuts}.
The reference CBCT image has a resolution of $256\times256\times256$ and a voxel size of $0.1961\mathrm{mm}\times0.1961\mathrm{mm}\times0.1961\mathrm{mm}$.
Uniformly spaced 5, 10, 20 X-ray projections from middle positioned scan are used as model input as mentioned in Sec.~\ref{sec:dataacq-realworld}.
The source-to-object distance $L^{sb}=\mathrm{66mm}$, and the object-to-detector distance $L^{bd}=\mathrm{133mm}$.}

\subsubsection{Implementation Details}
We adopt ResNet34~\cite{resnet} as the backbone of the 2D CNN encoder. 
The encoder consists of an initial CNN layer followed by four subsequent residual layers. 
For each query point, we concatenate the feature vectors interpolated from each layer's feature map to create the output feature vector. 
\lzt{As for the 3D CNN decoder,} we adopt the generator network structure from SRGAN~\cite{SRGAN,3DSRGAN}. 
This decoder is composed of six residual blocks at the beginning and a variable number of upsampling blocks. Each upsampling block increases the 3D feature map resolution by a factor of two. For instance, when $S=4$, there are two upsampling blocks. 
And we would include an additional upsampling block if $S$ doubled.

In our experiments, we empirically set $\lambda_{grad}=1$, $\lambda_{proj}=0.01$, downsampling rate $S=4$, channel size $C=256$, random ray batch size $\mathbf{\left|B\right|}=1024$.
We employ the Adam optimizer with an initial learning rate of 1$\times10^{-4}$, which decays by a factor of 0.5 every 50 epochs. We train the model for 200 epochs with a training batch size of 1. All experiments are conducted on a single A100 GPU. 

\subsubsection{Competing Methods and Evaluation Metrics} 

In our study, we evaluate our framework against a diverse set of methonds,
including conventional techniques like FDK and SART, 
\lzt{neural rendering-based approaches like NAF, SCOPE3D~\cite{SCOPE}, \R{and} SNAF, deep learning-based approaches like PatRecon, PixelNeRF, DIF-Net, and DDS3D~\cite{DDS}.
FDK and SART are implemented based on ASTRA-toolbox~\cite{astratoolbox}.
The other methods are directly adopted from their source codes, except the following ones.
SCOPE~\cite{SCOPE} incorporates hash-based neural rendering technique and re-projection strategy for sparse-view 2D CT reconstruction specified for parallel/fan beam geometry.
We have extended SCOPE to a cone beam setting for CBCT reconstruction based on NAF codebase, named SCOPE3D.
The primary difference between SCOPE3D and NAF is that SCOPE3D employs a re-projection strategy on its well-trained model to synthesize dense-view projections. 
Following that, it replaces the projections at the corresponding views in the synthesized dense-view projections with the given sparse input.
These swapped projections are then used for reconstruction with traditional methods such as FDK.
In our experiments, we use SART for the re-projection reconstruction instead.
PixelNeRF is originally designed for natural scene, we modify its forward projection equation to make it suitable for X-ray imaging.
DDS adopts 2D pretrained diffusion model to generate 3D CT image slice by slice.
It employs parallel beam forward projection and $z$-axis TV regularization to guide its inference sampling process.
We modify it to use cone beam forward projection for CBCT reconstruction, named DDS3D.}
To assess performance, we utilize the Peak Signal-to-Noise Ratio (PSNR) and Structural Similarity Index (SSIM)~\cite{SSIM} as our primary metrics.
\lzt{Furthermore, the efficiency analysis in terms of both time and memory would be presented in Sec.~\ref{sec:efficiency}.}

\subsection{Reconstruction Results}
\label{sec:recon_results}

\subsubsection{Qualitative Results}

\begin{figure*}[h!]
\centering
    \includegraphics[width=1.0\textwidth]{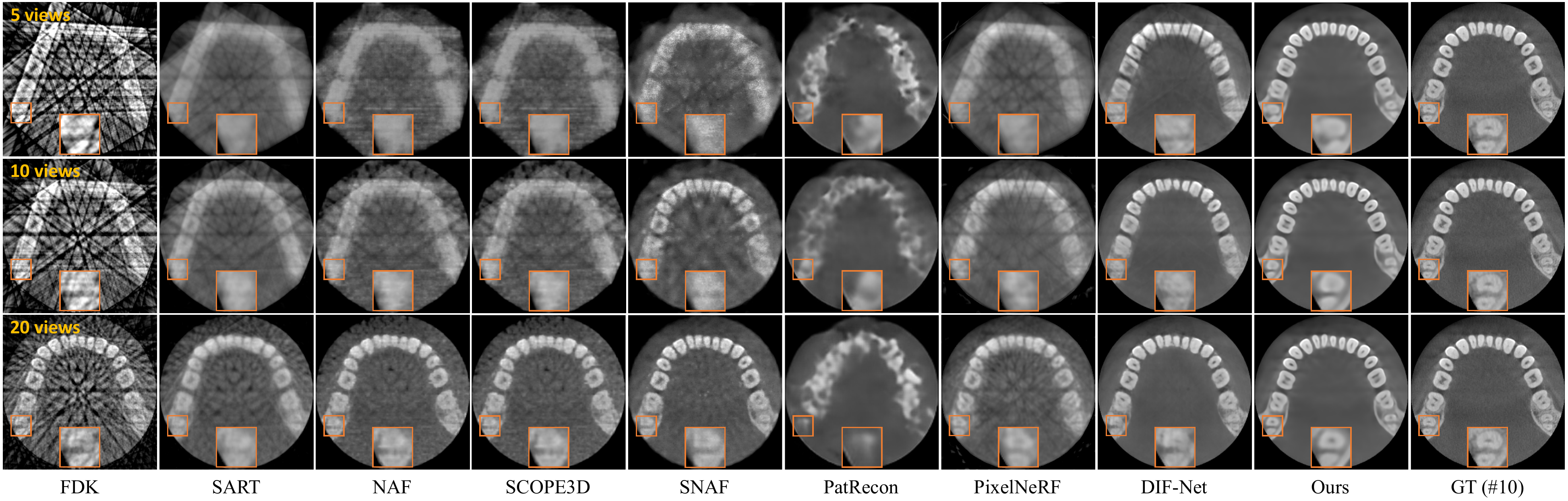}
    \caption{\lzt{Qualitative comparison on case \#10 from dental dataset (axial slice). Window: [-1000, 2000] HU.}
    } 
    \label{qualitative_case10_dental}
\end{figure*}

\begin{figure*}[h!]
\centering
    \includegraphics[width=1.0\textwidth]{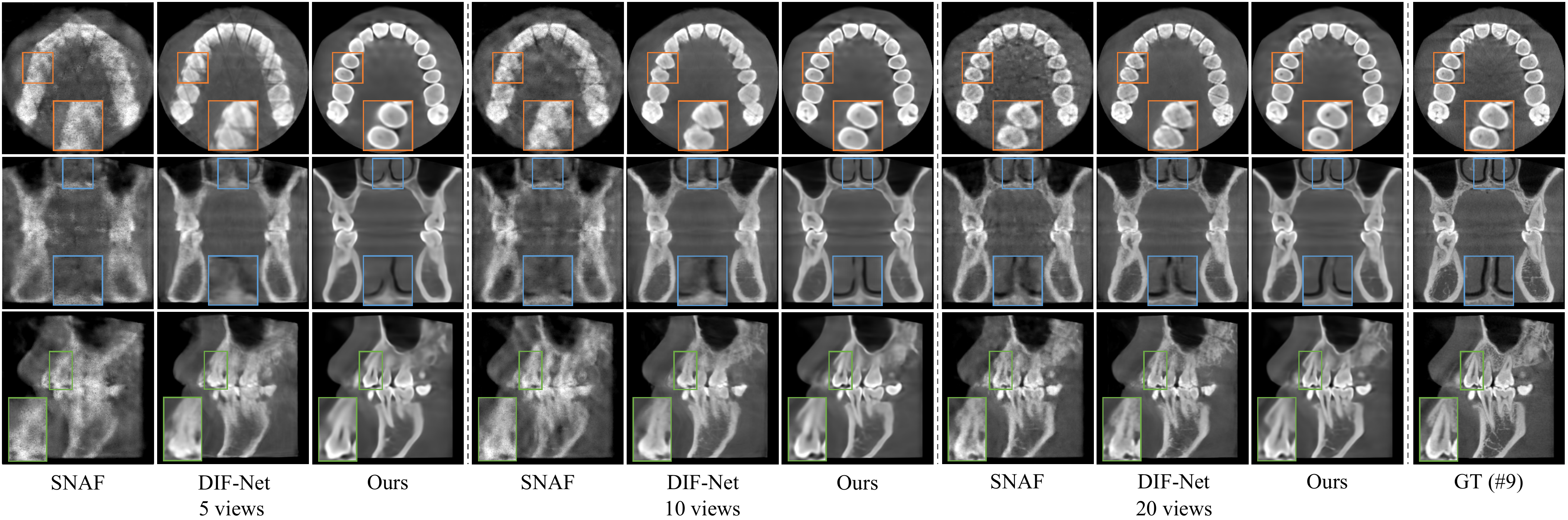}
    \caption{
    \lzt{Qualitative comparison with SNAF and DIF-Net on case \#9 from dental dataset. From top to bottom: axial, coronal, and sagittal slices. Window: [-1000, 2000] HU.}
    } 
    \label{three_axes_comparison_case9_dental}
\end{figure*}

\begin{figure*}[h!]
\centering
    \includegraphics[width=1.0\textwidth]{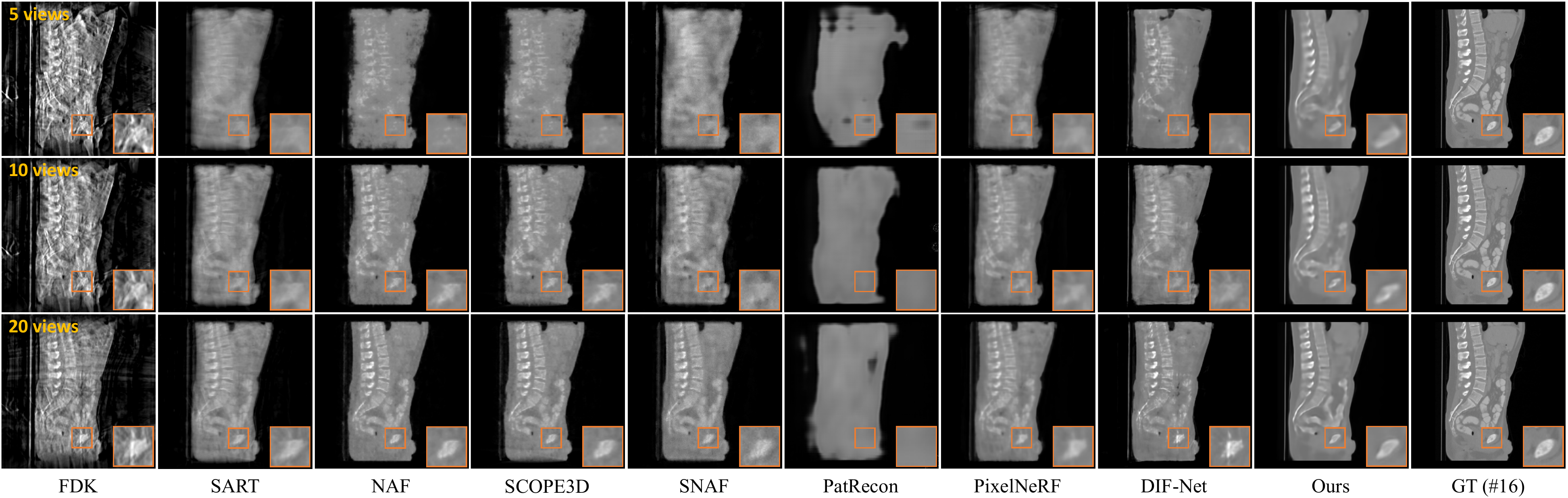}
    \caption{\lzt{Qualitative comparison on case \#16 from spine dataset (sagittal slice). Window: [-1000, 1000] HU.} 
    } 
    \label{qualitative_case16_spine}
\end{figure*}

\begin{figure*}[h!]
\centering
    \includegraphics[width=1.0\textwidth]{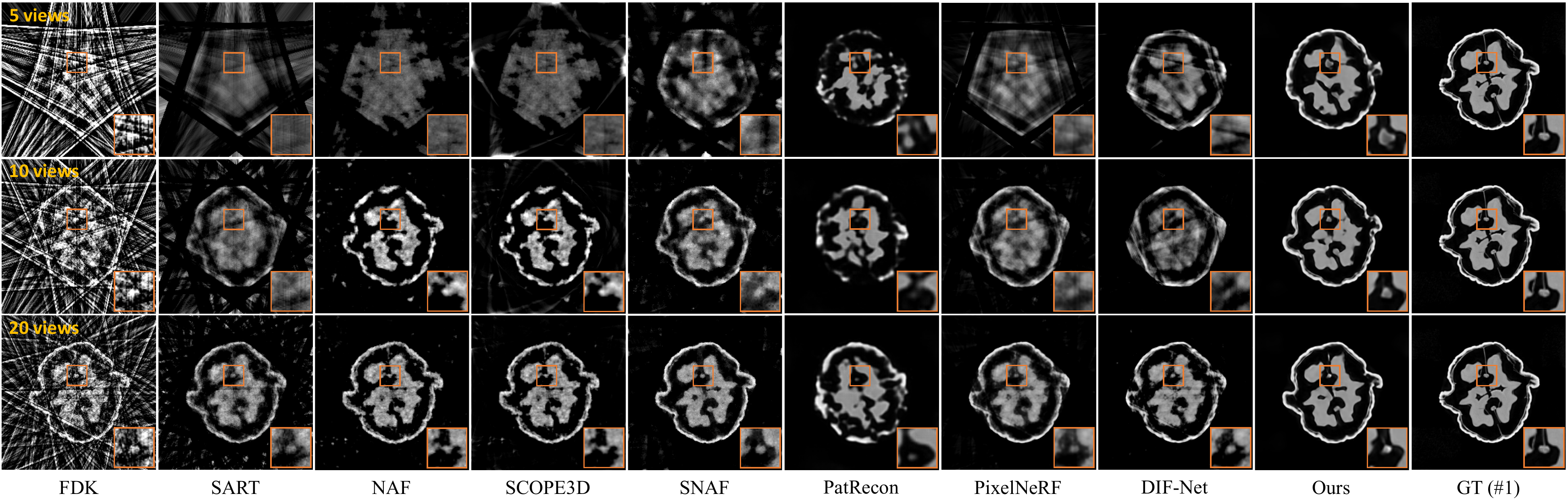}
    \caption{\lzt{Qualitative comparison on case \#1 from walnut dataset (axial slice). Window: [-1000, 2000] HU.} 
    } 
    \label{qualitative_case1_walnut}
\end{figure*}

Fig.~\ref{qualitative_case10_dental} presents a side-by-side comparison of 3D CBCT images reconstructed in axial slices from \lzt{case \#10 of} dental dataset. 
It is evident that the FDK approach struggles with sparse-view input, leading to significant streaky artifacts because of the limited number of input views. 
Although SART reduces these artifacts, it often loses fine details in the process. 
When it comes to neural rendering-based methods, NAF achieves \lzt{decent} results \lzt{with 20 views} by incorporating neural rendering with hash encoding. 
Yet, its performance greatly diminishes with very few input views (such as 5 or 10), as it is optimized for individual objects and lacks prior knowledge learned from the data population.
\lzt{SCOPE3D shows similar performance, as the re-projection strategy offers negligible new information.
SNAF demonsrates improvements due to its view augmentation strategy but still struggles with 10 or 5 views.}
PatRecon ignores geometric relationships among multi-view projections, which results in blurry reconstructions with \lzt{erroneous structures}.
\lzt{Benefiting from CNN layers, PixelNeRF enjoys prior knowledge and maintains multi-view consistency.
But it tends to produce noticeable streaky artifacts due to its point-wise MLP decoding and 2D supervision.}
\lzt{DIF-Net builds upon the principles of PixelNeRF, achieving better results due to its 3D supervision.
The results with 20 views input are comparable to ours, with slight blurriness and noise as highlighted in the orange box.
However, its performance degrades with sparser inputs, such as 5 views, exhibiting streaky artifacts due to its point-wise MLP decoding approach.}
\lzt{This is because the point-wise MLP independently decodes the attenuation of each query point, disregarding the spatial relationships among neighboring voxel points in CBCT image.
MLP decoder is also unable to capture the global structure information of CBCT image with the point-wise 3D supervision.
As a result, it would deliver streaky artifacts, especially when facing extremely sparse input like 5 views.
In contrast, our CNN-based decoding module considers interactions among neighboring points, effectively acting as a learnable filter to mitigate noise and extract more robust feature representations.
Moreover, 3D CNN decoder is capable of capturing the global structure information with the volume-wise 3D supervision.
Consequently, our reconstructed CBCT images exhibit higher quality with less streaky artifacts.}
\lzt{Notably, our approach surpasses all other methods, providing reconstruction quality comparable to the ground truth with 20 input views.
However, recovering details with high fidelity becomes challenging for our method when facing 10 or 5 views.
Despite this limitation, our method still maintains a clear advantage over the competition, showing less streaky artifacts and preserving a better global structure.}

\lzt{Additionally, we provide a detailed comparison with two current state-of-the-art methods, SNAF and DIF-Net, using case \#9 from the dental dataset.
This comparison includes axial, coronal, and sagittal slices, as shown in Fig.~\ref{three_axes_comparison_case9_dental}. 
It is clear that SNAF performs the worst, especially when limited to 10 or 5 views.
This is because its lack of leveraging prior knowledge from data populations, unlike DIF-Net and our method. 
DIF-Net shows some noises and streaky artifacts, particularly with 5 views, and its detail recovery is not as good as ours. 
Overall, our method achieves superior performance compared to the others. 
However, we also acknowledge that there can be some blurriness and minor structural errors with our method when limited to 10 or 5 views.}

Fig.~\ref{qualitative_case16_spine} provides visual comparison of reconstructed images in sagittal slices \lzt{using case \#16} from spinal dataset. 
We notice a consistent trend, with our method outperforming the others in performance. 
However, a common challenge emerges across all methods, including ours, in accurately reconstructing soft tissues and organs in the abdomen. 
This difficulty is due to the nature of spinal scans, which encompass a variety of organs and soft tissues. 
Many of these soft tissues exhibit low contrast differences, making it hard to achieve clear reconstructions with limited X-ray views.
In comparison, dental scans primarily focus on teeth and jawbones, which present more distinct contrasts. 
Therefore, reconstructing sparse-view CBCT images is particularly challenging for spinal scans due to their intricate content. 
Despite these hurdles, the successful application of our method to both dental and spinal datasets showcases its versatility and robustness across different body parts.

\lzt{Fig.~\ref{qualitative_case1_walnut} provides visual comparison of reconstructed CBCT image in axial slices using case \#1 from walnut dataset.
When dealing with 20 views input, SNAF and our method stand out with satisfactory results, albeit with some missing details.
When reduced to 10 or 5 views, all methods, including ours, exhibit structural errors and do not perform well.
Our method's results are relatively clear, with less noise and streaky artifacts, and the overall structure and quality are better compared to other methods.
We have to admit that the walnut dataset is particularly difficult for reconstruction.
This is partially due to the presence of noises from real-world captures.
More importantly, the extremely low contrast between different regions within the walnut image poses another significant challenge.
Despite such challenges, our method successfully adapts to real-world walnut dataset, further verifying our robustness.
Based on the results from both the spine and walnut datasets, improving reconstruction quality for low-contrast images remains a challenging yet meaningful research direction.}

\subsubsection{Quantitative Results}

\begin{table}[h!]
\caption{\lzt{Quantitative comparison on dental dataset. The best performance is shown in bold.}
}
\centering
\fontsize{5}{6}\selectfont
\setlength{\tabcolsep}{5pt}
\renewcommand\arraystretch{1.62}
\begin{tabular}{c|cc|cc|cc}
\hline
\multirow{2}{*}{Method} & \multicolumn{2}{c|}{5 views}     & \multicolumn{2}{c|}{10 views}    & \multicolumn{2}{c}{20 views}     \\ \cline{2-7} 
                        & PSNR          & SSIM            & PSNR           & SSIM            & PSNR           & SSIM            \\ \hline
FDK                     & 16.31$\pm$0.43 & 0.221$\pm$0.014 & 18.53$\pm$0.47 & 0.301$\pm$0.017 & 22.56$\pm$0.56 & 0.422$\pm$0.021 \\
SART                    & 20.22$\pm$1.82 & 0.621$\pm$0.024 & 24.17$\pm$1.13 & 0.704$\pm$0.018 & 27.93$\pm$0.91 & 0.784$\pm$0.015 \\
NAF                     & 21.55$\pm$1.21 & 0.578$\pm$0.036 & 23.89$\pm$1.14 & 0.673$\pm$0.024 & 28.77$\pm$0.86 & 0.793$\pm$0.020 \\
SCOPE3D                 & 21.74$\pm$0.99 & 0.588$\pm$0.024 & 24.05$\pm$1.12 & 0.684$\pm$0.023 & 29.39$\pm$0.87 & 0.807$\pm$0.019 \\
SNAF                    & 23.46$\pm$0.47 & 0.608$\pm$0.023 & 25.97$\pm$0.51 & 0.706$\pm$0.019 & 30.93$\pm$0.51 & 0.844$\pm$0.015 \\
PatRecon                & 19.89$\pm$0.89 & 0.573$\pm$0.046 & 19.91$\pm$0.66 & 0.574$\pm$0.030 & 19.95$\pm$0.85 & 0.569$\pm$0.038 \\
PixelNeRF               & 22.12$\pm$1.36 & 0.643$\pm$0.023 & 24.03$\pm$0.91 & 0.710$\pm$0.018 & 26.85$\pm$0.57 & 0.775$\pm$0.014 \\
DIF-Net                 & 25.80$\pm$0.93 & 0.759$\pm$0.027 & 27.52$\pm$0.84 & 0.818$\pm$0.021 & 30.48$\pm$0.80 & 0.870$\pm$0.015 \\
Ours                    & \textbf{27.48$\pm$1.12} & \textbf{0.823$\pm$0.028} & \textbf{28.83$\pm$1.19} & \textbf{0.850$\pm$0.025} & \textbf{31.44$\pm$1.00} & \textbf{0.891$\pm$0.016} \\ \hline
\end{tabular}
\label{tab:compare_dental}
\end{table}

\begin{table}[h!]
\caption{\lzt{Quantitative comparison on spine dataset. The best performance is shown in bold.}
}
\centering
\fontsize{5}{6}\selectfont
\setlength{\tabcolsep}{5pt}
\renewcommand\arraystretch{1.61}
\begin{tabular}{c|cc|cc|cc}
\hline
\multirow{2}{*}{Method} & \multicolumn{2}{c|}{5 views}                       & \multicolumn{2}{c|}{10 views}                      & \multicolumn{2}{c}{20 views}                       \\ \cline{2-7} 
                        & PSNR                    & SSIM                     & PSNR                    & SSIM                     & PSNR                    & SSIM                     \\ \hline
FDK                     & 17.06$\pm$1.27          & 0.258$\pm$0.051          & 20.27$\pm$1.30          & 0.286$\pm$0.044          & 22.87$\pm$1.37          & 0.352$\pm$0.041          \\
SART                    & 19.65$\pm$1.96          & 0.743$\pm$0.067          & 22.60$\pm$2.27          & 0.794$\pm$0.056          & 27.08$\pm$1.96          & 0.846$\pm$0.039          \\
NAF                     & 20.58$\pm$2.92          & 0.781$\pm$0.063          & 26.02$\pm$2.50          & 0.861$\pm$0.036          & 30.80$\pm$1.93          & 0.912$\pm$0.024          \\
SCOPE3D                 & 20.77$\pm$2.78          & 0.779$\pm$0.062          & 26.24$\pm$2.38          & 0.858$\pm$0.035          & 30.91$\pm$2.00          & 0.908$\pm$0.024          \\
SNAF                    & 22.05$\pm$1.48          & 0.711$\pm$0.075          & 26.33$\pm$1.85          & 0.806$\pm$0.062          & 31.69$\pm$1.39          & 0.902$\pm$0.035          \\
PatRecon                & 18.30$\pm$1.62          & 0.686$\pm$0.054          & 18.29$\pm$2.14          & 0.681$\pm$0.061          & 18.57$\pm$1.75          & 0.641$\pm$0.047          \\
PixelNeRF               & 20.81$\pm$1.98          & 0.772$\pm$0.075          & 25.41$\pm$1.67          & 0.848$\pm$0.043          & 28.68$\pm$1.64          & 0.890$\pm$0.033          \\
DIF-Net                 & 25.75$\pm$2.46          & 0.825$\pm$0.061          & 28.65$\pm$2.00          & 0.858$\pm$0.052          & 32.28$\pm$1.75          & 0.901$\pm$0.034          \\
DDS3D                   & 22.31$\pm$1.62          & 0.405$\pm$0.083          & 23.94$\pm$2.52          & 0.505$\pm$0.102          & 26.25$\pm$1.33         & 0.602$\pm$0.072          \\
Ours                    & \textbf{28.32$\pm$1.80} & \textbf{0.884$\pm$0.035} & \textbf{31.50$\pm$1.94} & \textbf{0.921$\pm$0.025} & \textbf{33.14$\pm$1.88} & \textbf{0.938$\pm$0.020} \\ \hline
\end{tabular}
\label{tab:compare_spine}
\end{table}

\begin{table}[h!]
\caption{\lzt{Quantitative comparison on walnut dataset. The best performance is shown in bold.}
}
\centering
\fontsize{5}{6}\selectfont
\setlength{\tabcolsep}{5pt}
\renewcommand\arraystretch{1.61}
\begin{tabular}{c|cc|cc|cc}
\hline
\multirow{2}{*}{Method} & \multicolumn{2}{c|}{5 views}      & \multicolumn{2}{c|}{10 views}    & \multicolumn{2}{c}{20 views}     \\ \cline{2-7} 
                        & PSNR           & SSIM           & PSNR           & SSIM            & PSNR           & SSIM            \\ \hline
FDK                     & 11.08$\pm$0.18  & 0.114$\pm$0.003 & 13.25$\pm$0.18 & 0.141$\pm$0.004 & 16.09$\pm$0.19 & 0.191$\pm$0.003 \\
SART                   & 17.25$\pm$0.29  & 0.378$\pm$0.013 & 19.55$\pm$0.29 & 0.487$\pm$0.016 & 23.71$\pm$0.40 & 0.588$\pm$0.012 \\
NAF                    & 18.15$\pm$1.17  & 0.522$\pm$0.047 & 21.72$\pm$0.82 & 0.612$\pm$0.027 & 25.36$\pm$0.22 & 0.680$\pm$0.006 \\
SCOPE3D                 & 19.08$\pm$0.64 & 0.523$\pm$0.046 & 20.75$\pm$1.27 & 0.535$\pm$0.046 & 25.92$\pm$0.19 & 0.714$\pm$0.009 \\
SNAF                    & 19.51$\pm$0.33  & 0.542$\pm$0.015 & 23.29$\pm$0.37 & 0.651$\pm$0.009 & \textbf{27.43$\pm$0.29} & 0.748$\pm$0.006\\
PatRecon                & 17.54$\pm$0.41  & 0.713$\pm$0.024 & 17.51$\pm$0.43 & 0.712$\pm$0.022 & 17.30$\pm$0.55 & 0.699$\pm$0.033 \\
PixelNeRF               & 19.87$\pm$0.19  & 0.482$\pm$0.009 & 23.41$\pm$0.43 & 0.682$\pm$0.009 & 24.79$\pm$0.15 & 0.732$\pm$0.004 \\
DIF-Net                 & 21.30$\pm$0.34  & 0.640$\pm$0.006 & 21.76$\pm$0.45 & 0.640$\pm$0.008 & 25.99$\pm$0.14 & 0.707$\pm$0.004 \\
Ours                    & \textbf{21.64$\pm$0.53}  & \textbf{0.811$\pm$0.012} & \textbf{24.78$\pm$0.35} & \textbf{0.867$\pm$0.010} & 26.84$\pm$0.37 & \textbf{0.895$\pm$0.008} \\ \hline
\end{tabular}
\label{tab:compare_walnut}
\end{table}

\lzt{To evaluate our method statistically, we present quantitative comparisons in Tab.~\ref{tab:compare_dental}, Tab.~\ref{tab:compare_spine}, and Tab.~\ref{tab:compare_walnut} for the dental, spine, walnut dataset, respectively.
The performance trend is similar to qualitative results.
For the dental and spine datasets, our method consistently outperforms others, as measured by PSNR and SSIM metrics.
On the walnut dataset, our method achieves the best performance for both the 10 views and 5 views input conditions.
Under the 20 views input condition, our method ranks second in PSNR, slightly lower than SNAF, but achieves the highest SSIM. 
A possible explanation is that methods based on per-scene optimization may be more robust to noisy real-world data. 
The noises vary among different cases, which may hinder the learning of generalization methods.
Another reason is that the available training data is too limited on walnut dataset (32 cases), preventing our method from fully extracting prior knowledge from such a small data population.
Due to these two factors, our method's PSNR metric is slightly lower than that of SNAF.
Overall, considering all datasets, all view settings, and two metrics, our method achieves the best performance.}

\subsubsection{Compare with Diffusion Model}
\label{sec:compare_with_diffusion}

\begin{figure}[]
\centering
    \includegraphics[width=0.5\textwidth]{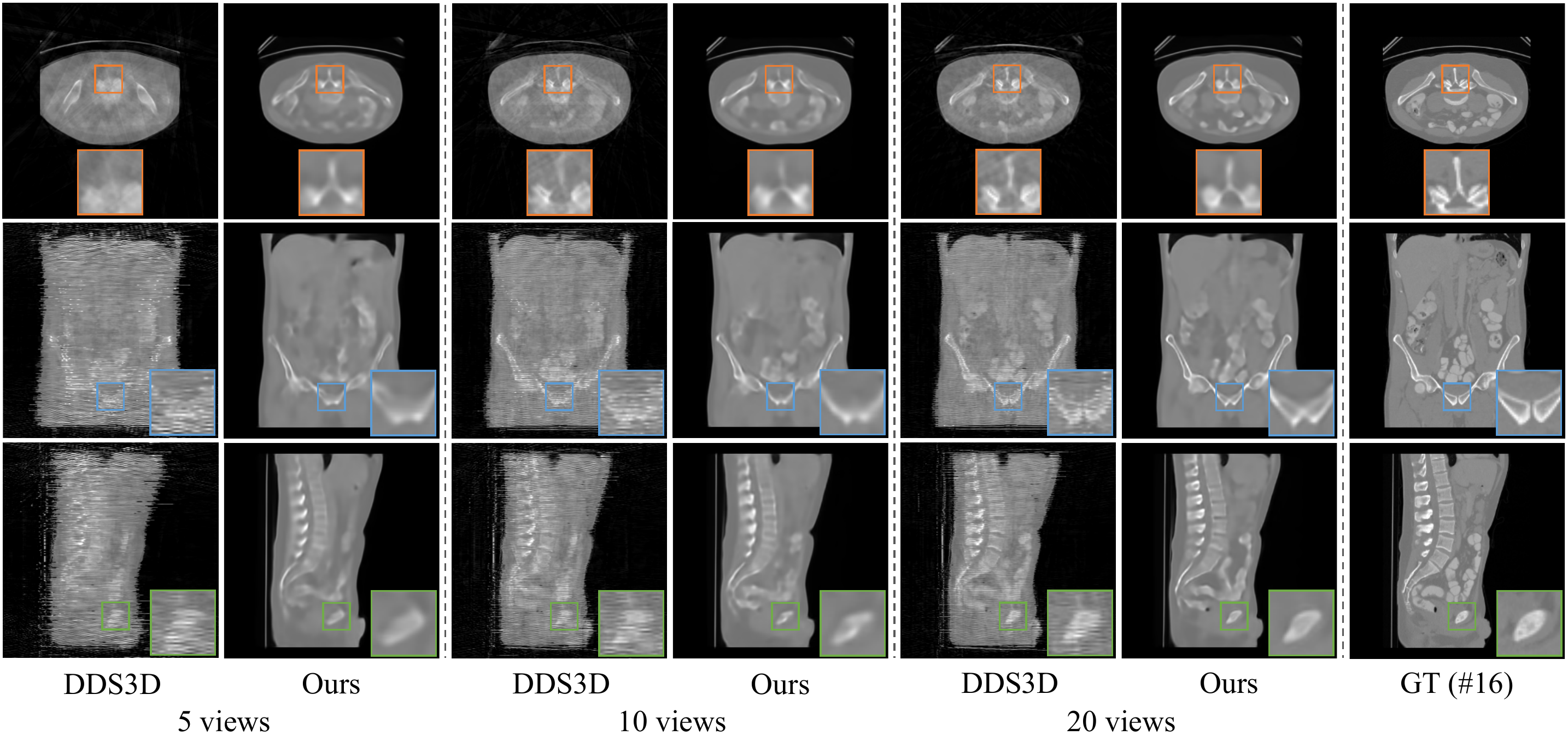}
    \caption{
    \lzt{Qualitative comparison with DDS3D on case \#16 from spine dataset. From top to bottom: axial, coronal, and sagittal slices. Window: [-1000, 1000] HU.}
    } 
    \label{fig:diffusion comparison}
    \vspace{-4mm}
\end{figure}

\begin{table}[]
\caption{\lzt{Efficiency analysis. The best-performing is shown in bold. Unit: Time (s); Mem. (GB): GPU Memory consumption in training; Size (MB).}}
\scriptsize
\centering
\setlength{\tabcolsep}{4pt}
\renewcommand\arraystretch{1.5}
\begin{tabular}{c|cc|cc|cc|c}
\hline
\multirow{2}{*}{Method} & \multicolumn{2}{c|}{5 views} & \multicolumn{2}{c|}{10 views} & \multicolumn{2}{c|}{20 views} & \multirow{2}{*}{Size} \\ \cline{2-7}
                        & Time           & Mem.        & Time           & Mem.         & Time           & Mem.         &                       \\ \hline
FDK                     & 0.48           & -           & 0.51           & -            & 0.50           & -            & -                     \\
SART                    & 102.15         & -           & 127.18         & -            & 111.65         & -            & -                     \\
NAF                     & 212.29         & 12.74       & 403.38         & 12.74        & 787.26         & 12.74        & 54.28                 \\
SCOPE3D                 & 718.46         & 12.74       & 933.47         & 12.74        & 1303.97        & 12.74        & 54.28                 \\
SNAF                    & 1681.20        & 7.03        & 1685.98        & \textbf{7.03}         & 1806.09        & \textbf{7.03}         & 134.83                \\
PatRecon                & \textbf{0.007}          & 21.25       & \textbf{0.007}          & 21.25        & \textbf{0.012}          & 21.25        & 1557.40                \\
PixelNeRF               & 5.38           & 17.35       & 7.84           & 24.56        & 12.77          & 35.39        & \textbf{26.35}                 \\
DIF-Net                 & 1.60           & \textbf{5.05}        & 3.35           & 7.82         & 7.62           & 14.45        & 118.66                \\
DDS3D                   & 1073.11        & 8.83        & 2072.44        & 8.83         & 4218.67        & 8.97         & 108.51                \\
Ours                    & 0.53           & 32.86       & 0.65           & 39.95        & 0.93           & 68.69        & 104.83                \\ \hline
\end{tabular}
\label{tab: efficiency}
\vspace{-4mm}
\end{table}

\lzt{In this section, we compare our method with diffusion-based method DDS3D on spine dataset.
The quantitative results are presented in Tab.~\ref{tab:compare_spine}, and a visual example is shown in Fig.~\ref{fig:diffusion comparison}.
It is clear that our method outperforms DDS3D by a large margin in both PSNR and SSIM metrics as well as in visual quality.
When dealing with 20 views, DDS3D provides decent result on axial slices, although there are still some streaky artifacts and noises.
If we observe the coronal and sagittal slices, we find that DDS3D exhibits severe inter-slice inconsistency due to its slice-wise diffusion sampling process. 
This inconsistency is particularly noticeable at the upper and lower ends, attributed to cone beam forward projection guided sampling. 
The sparse nature of cone beam ray casting at these ends leads to insufficient supervisory signals.
This issue becomes more pronounced with fewer input views, consequently degrading the reconstruction quality.
Overall, despite DDS~\cite{DDS} performing well in CT reconstruction with parallel beam geometry, its performance with cone beam geometry is still far from satisfactory.}

\subsubsection{Efficiency Analysis}
\label{sec:efficiency}

\lzt{We further evaluate each method’s efficiency in terms of both time and memory. 
Time efficiency includes reconstruction time for one case, while memory efficiency encompasses model size and GPU memory consumption during the training stage. 
The quantitative results are presented in Tab.~\ref{tab: efficiency}.
Note that the metrics of SNAF are reported on its optimization stage, and those for DDS3D are reported on its diffusion sampling process.} 

\lzt{Our method could complete reconstruction within a second, demonstrating exceptional time efficiency. 
It is much faster than individual optimization methods (e.g., SART, NAF, SCOPE3D, SNAF) and diffusion-based method (e.g., DDS3D), which are hindered by time-consuming iterative calculation and slice-wise diffusion sampling process, respectively. 
PixelNeRF and DIF-Net, while also time efficient, face delays because they need to query full-resolution features from different views during reconstruction. 
In contrast, our method samples query points at a lower resolution, making it even faster. 
What’s more, our model size is also manageable compared with other methods. 
However, it is important to note that our model is not GPU memory efficient, as it requires considerable GPU memory during the training phase. 
And it will be included as one of our limitations.}

\subsection{Ablation Study}
\label{sec:ablation}

\lzt{In this section, we study the influence of our different model components, including feature fusing strategy, loss term, loss weight, and downsampling rate.
Note that, our model is trained on dental dataset under the setting of downsampling rate $S=4$, 20 input views, and uniformly sampled angles within the range of $[0^\circ, 360^\circ)$, unless otherwise specified.}

\subsubsection{Feature Fusing Strategy}
\label{sec:ab_on_fusing}

\begin{table}[]
\caption{\lzt{Quantitative results of ablation study on feature fusing strategy on dental dataset.}}
\centering
\setlength{\tabcolsep}{5pt}
\renewcommand\arraystretch{1.5}
\begin{tabular}{c|cc}
\hline
Fusing                                    & PSNR             & SSIM            \\ \hline
Max                                        & 29.87$\pm$1.17 & 0.870$\pm$0.020 \\
Average                                    & 30.74$\pm$1.34 & 0.883$\pm$0.017 \\ \hline
$\mathbf{f}_{i}$                            & 30.70$\pm$0.76 & 0.881$\pm$0.015 \\
$\mathbf{f}_{i} \oplus \mathbf{f}_{\sigma}$ & 31.19$\pm$0.86 & 0.884$\pm$0.015 \\
$\mathbf{f}_{i} \oplus \mathbf{f}_{\mu}$    & 31.27$\pm$0.96 & 0.889$\pm$0.015 \\ \hline
Adaptive                                   & 31.44$\pm$1.00 & 0.891$\pm$0.016 \\ \hline
\end{tabular}
\label{tab_ab_on_fusion}
\vspace{-4mm}
\end{table}

\begin{figure*}[]
\centering
    \includegraphics[width=0.85\textwidth]{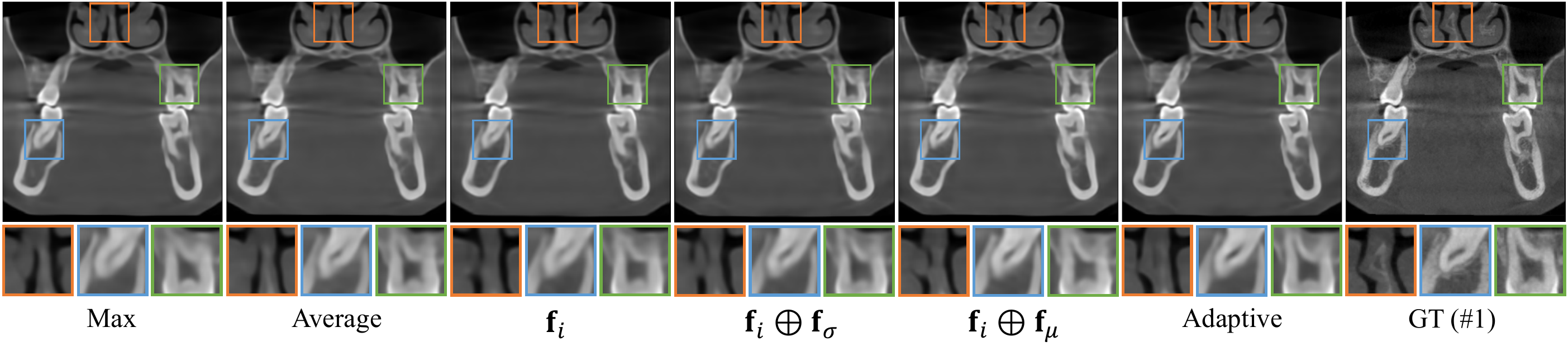}
    \caption{
        \lzt{Qualitative results of ablation study on feature fusing strategy on case \#1 from dental dataset (coronal slice). Window: [-1000, 2000] HU.}
    }
    \label{fig: fusing strategy}
    \vspace{-2.5mm}
\end{figure*}

We first study the influence of different feature fusing strategies, including max pooling, average pooling, and our designed adaptive pooling.
\lzt{Besides, we ablate each component $\mathbf{f}_i$, $\mathbf{f}_\mu$, and $\mathbf{f}_\sigma$ in our adaptive fusion module to study their importance.
The quantitative results are delivered in Tab.~\ref{tab_ab_on_fusion}, and a visual example is shown in Fig.~\ref{fig: fusing strategy}. 
Take the setting of $\mathbf{f}_i$ for example, the input of the first MLP block $\phi_1$ in Eq.~\ref{firstmlpfusion} is replaced by $\mathbf{f}_i$.}

\lzt{It's obvious that max pooling performs the worst due to significant information loss.
Average pooling retains all the information from different views, leading to performance improvement.
However, the feature from each view lacks interaction, limiting performance gains.
For instance, the low-contrast regions highlighted in Fig.~\ref{fig: fusing strategy}, such as the nasal cavity and tooth pulp, remain poorly reconstructed.
While the setting of $\mathbf{f}_i$ also retains information from various views, it still lacks interaction among different views.
Both of $\mathbf{f}_i \oplus \mathbf{f}_{\sigma}$ and $\mathbf{f}_i \oplus \mathbf{f}_{\mu}$ incorporate global feature term $\mathbf{f}_{\sigma}$ or $\mathbf{f}_{\mu}$, enabling $\mathbf{f}_{i}$ to interact with features from other views through the first MLP block $\phi_1$, resulting in more effective feature fusion and enhanced performance.
Moreover, incorporating $\mathbf{f}_{\mu}$ yields a slight better performance than $\mathbf{f}_{\sigma}$, indicating higher importance of $\mathbf{f}_{\mu}$.}
\lzt{Finally, our full adaptive feature fusing design achieves the best results, with visual details closely matching the ground truth and yielding the highest metrics.}
Overall, the adaptive feature fusion strategy demonstrates superior reconstruction performance, capturing finer details, especially in regions with low contrast.

\subsubsection{Loss Term}
\label{sec:ab_on_loss}

\begin{table}[]
 \caption{\lzt{Quantitative results of ablation study on loss term on dental dataset.}}
\centering
\setlength{\tabcolsep}{5pt}
\renewcommand\arraystretch{1.5}
\begin{tabular}{ccc|cc}
\hline
$L_{recon}$  & $L_{grad}$   & $L_{proj}$   & PSNR           & SSIM           \\ \hline
$\checkmark$ &              &              & 30.33$\pm$1.05 & 0.870$\pm$0.016 \\
$\checkmark$ & $\checkmark$ &              & 30.93$\pm$0.94 & 0.887$\pm$0.016 \\
$\checkmark$ & $\checkmark$ & $\checkmark$ & 31.44$\pm$1.00 & 0.891$\pm$0.016 \\ \hline
\end{tabular}
\label{tab: ab_on_loss}
\vspace{-2mm}
\end{table}

\begin{figure}[]
\centering
    \includegraphics[width=0.49\textwidth]{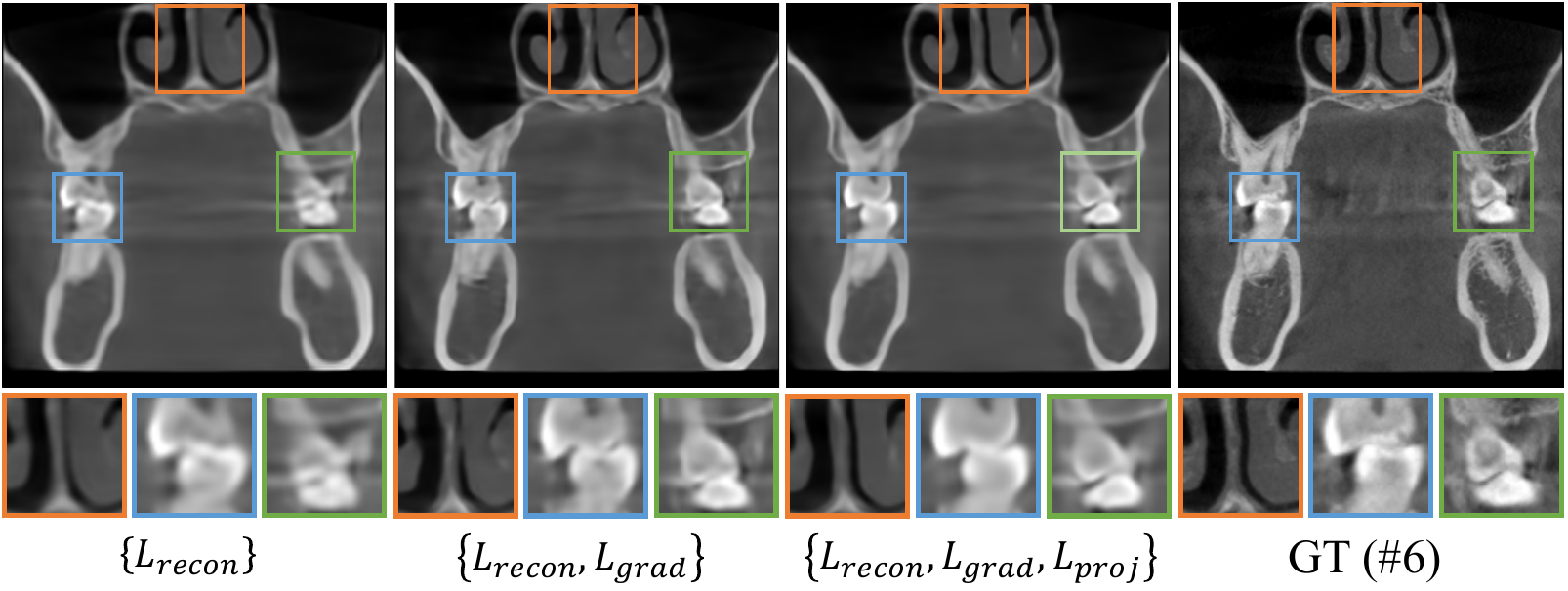}
    \caption{
        \lzt{Qualitative results of ablation study on loss term on case \#6 from dental dataset (coronal slice). Window: [-1000, 2000] HU.}
    } 
    \label{fig: ab_on_loss}
    \vspace{-3mm}
\end{figure}

Second, we also study the effectiveness of our different loss terms, including the 3D reconstruction loss ($L_{recon}$), gradient loss ($L_{grad}$), and 2D projection loss ($L_{proj}$).
Quantitative results are presented in Tab.~\ref{tab: ab_on_loss}.
We use $L_{recon}$ as our baseline, and each symbol $\checkmark$ indicates the inclusion of the particular loss term in the training \lzt{process}, thereby offering an alternative solution.
Our baseline model, using just the 3D reconstruction loss $L_{recon}$, already shows impressive performance, achieving over 30 $\mathrm{dB}$ for the dental dataset. 
This highlights the importance of 3D supervision.
With the addition of $L_{grad}$ and $L_{proj}$, the PSNR and SSIM values increase gradually.
A visual comparison is available in Fig.~\ref{fig: ab_on_loss}.
It is noticeable that $L_{grad}$ assists in recovering sharper details \lzt{around teeth pulp areas}.
However, it can also introduce some noise.
$L_{proj}$, acting as a regularization term, effectively reduces these noises and contributes to more accurate recovery of anatomical structures.

\subsubsection{Loss Weight}

\begin{figure}[]
\centering
    \includegraphics[width=0.5\textwidth]{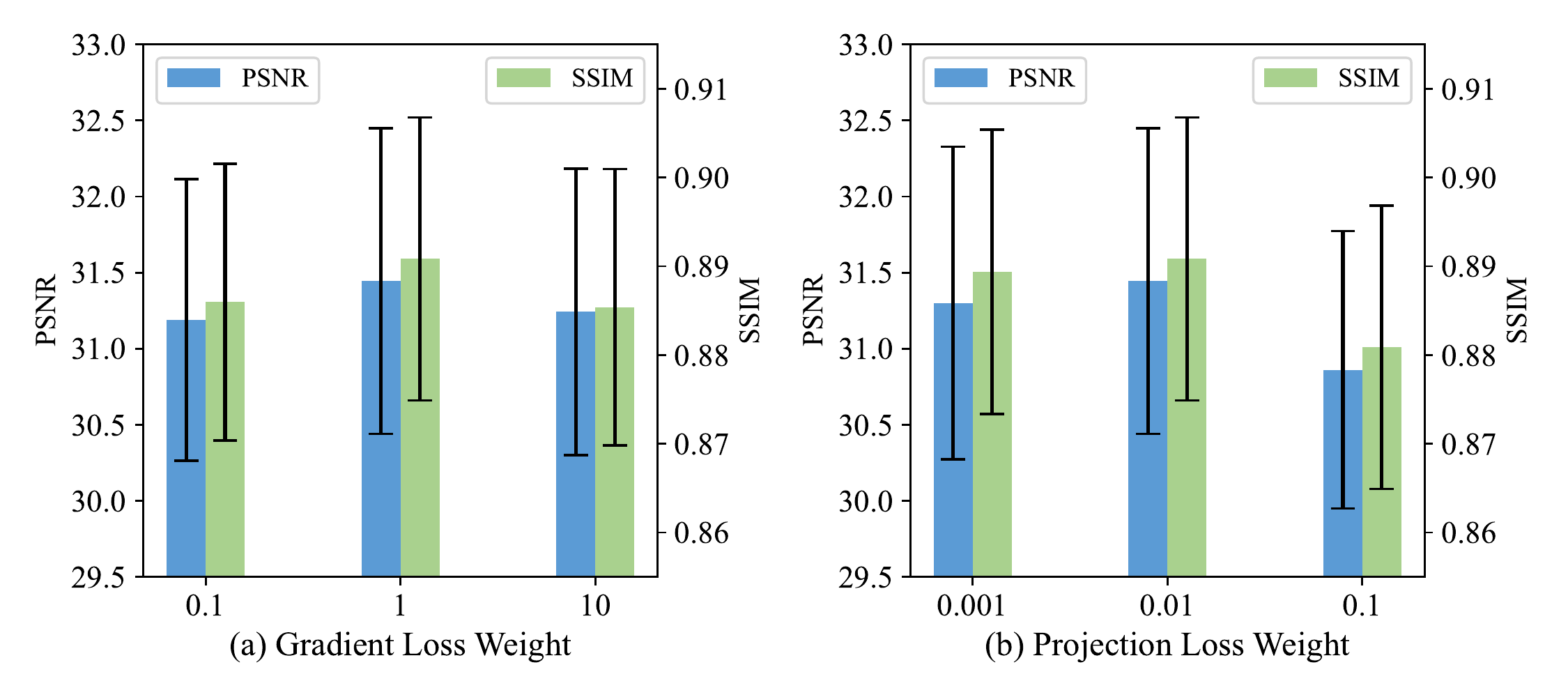}
    \caption{
        \lzt{Quantitative results of ablation study on (a) gradient loss weight and (b) projection loss weight on dental dataset.}
    } 
    \label{fig: ab_on_loss_weight}
    \vspace{-2mm}
\end{figure}

\begin{table}[]
\caption{\lzt{Ablation study on downsampling rate $S$. Unit: Time (s); Size (MB); Mem. (GB): GPU memory consumption in training.}}
\centering
\setlength{\tabcolsep}{6pt}
\renewcommand\arraystretch{1.5}
\begin{tabular}{cccccccc}
\hline
$N$                 & $S$ & PNSR        & SSIM             & Time & Size & Mem. \\ \hline
\multirow{3}{*}{5}  & 16  & 24.44 $\pm$ 0.71 & 0.748 $\pm$ 0.027 & 0.21     & 104.94     & 11.43     \\
                    & 8   & 26.35 $\pm$ 0.92 & 0.793 $\pm$ 0.028 & 0.35     & 104.88     & 13.85     \\
                    & 4   & 27.48 $\pm$ 0.71 & 0.823 $\pm$ 0.029 & 0.53     & 104.83     & 32.86     \\ \hline
\multirow{3}{*}{10} & 16  & 24.98 $\pm$ 0.75 & 0.760 $\pm$ 0.026 & 0.33     & 104.94     & 12.49     \\
                    & 8   & 27.21 $\pm$ 0.88 & 0.816 $\pm$ 0.026 & 0.43     & 104.88     & 15.91     \\
                    & 4   & 27.21 $\pm$ 0.88 & 0.842 $\pm$ 0.021 & 0.65     & 104.83     & 39.95     \\ \hline
\multirow{3}{*}{20} & 16  & 25.68 $\pm$ 0.67 & 0.776 $\pm$ 0.025 & 0.45     & 104.94     & 14.59     \\
                    & 8   & 28.66 $\pm$ 0.91 & 0.844 $\pm$ 0.021 & 0.90     & 104.88     & 19.47     \\
                    & 4   & 31.44 $\pm$ 1.00 & 0.891 $\pm$ 0.016 & 0.95     & 104.83     & 68.69     \\ \hline
\end{tabular}
\label{tab_ab_on_S}
\vspace{-4mm}
\end{table}

\lzt{Third, we verify our selection of the loss weights $\lambda_{grad}$ and $\lambda_{proj}$, which determine the relative importance of the gradient loss and projection loss, respectively.
Gradient loss helps model more effectively recover details, and projection loss enforces the model output aligns with the input} 
\lzt{X-ray projections.
In our work, we consider gradient loss is more important that projection loss.}

\lzt{Determining the optimal weights $\lambda_{grad}$ and $\lambda_{proj}$ through grid search can be experimentally laborious. 
Therefore, we perform a rough verification to support our choice.
The quantitative results with both mean and standard deviation are presented in Fig.~\ref{fig: ab_on_loss_weight}.
For the ablation study on the gradient loss weight $\lambda_{grad}$, we fixed $\lambda_{proj}=0.01$.
Conversely, for the ablation study on the projection loss weight $\lambda_{proj}$, we fixed $\lambda_{grad}=1$.
The results clearly demonstrate that our chosen values $\lambda_{grad}=1$ and $\lambda_{proj}=0.01$ yield the best performance in terms of both PSNR and SSIM metrics.}

\subsubsection{Downsampling Rate}

\lzt{At last}, we assess the effect of the downsampling rate $S$. 
The quantitative results are presented in Tab.~\ref{tab_ab_on_S}.
\lzt{Note that, $N$ represents the number of input views in this table.
The performance metrics (PSNR, SSIM) are reported on dental dataset.}
The trend is clear that, as the downsampling rate $S$ decreases, there is an improvement in reconstruction performance, as well as an increase in \lzt{reconstruction time}, and \lzt{GPU} memory usage \lzt{in training stage}.
\lzt{The model size slightly decreases for less unsampling blocks in decoder.}
A smaller $S$ means denser sampling on the voxel grid, which allows for capturing of more X-ray projection \lzt{information}, thereby enhancing the reconstruction quality. 
However, this benefit comes with the trade-off of higher computational demands and longer sampling \lzt{time}.
\lzt{$S=4$ offers the best performance among 5, 10, and 20 views, and the computation cost is also affordable.}
Therefore, we choose $S=4$ for our experiments.

\subsection{Robustness Analysis}

\lzt{In this section, we analyze our well-trained framework robustness to varying test conditions that differ from the training settings, including reconstruction resolution, angle sampling, and the number of input views. 
Note that, our model is trained on dental dataset under the setting of downsampling rate $S=4$, 20 input views, and uniformly sampled angles within the range of $[0^\circ, 360^\circ)$, unless otherwise specified.}

\subsubsection{Reconstruction Resolution}

\begin{table}[t]
\caption{\lzt{Quantitative results of robustness analysis on reconstruction resolution on dental dataset.}}
\centering
\setlength{\tabcolsep}{5pt}
\renewcommand\arraystretch{1.5}
\begin{tabular}{c|cc}
\hline
Test Resolution & PSNR  & SSIM    \\ \hline
128 res            & 24.85$\pm$0.43                & 0.783$\pm$0.017 \\
256 res            & 31.44$\pm$1.00                & 0.891$\pm$0.016 \\ \hline
\end{tabular}
\label{tab: ab recon res}
\vspace{-2mm}
\end{table}

\begin{figure}[t]
\centering
    \includegraphics[width=0.49\textwidth]{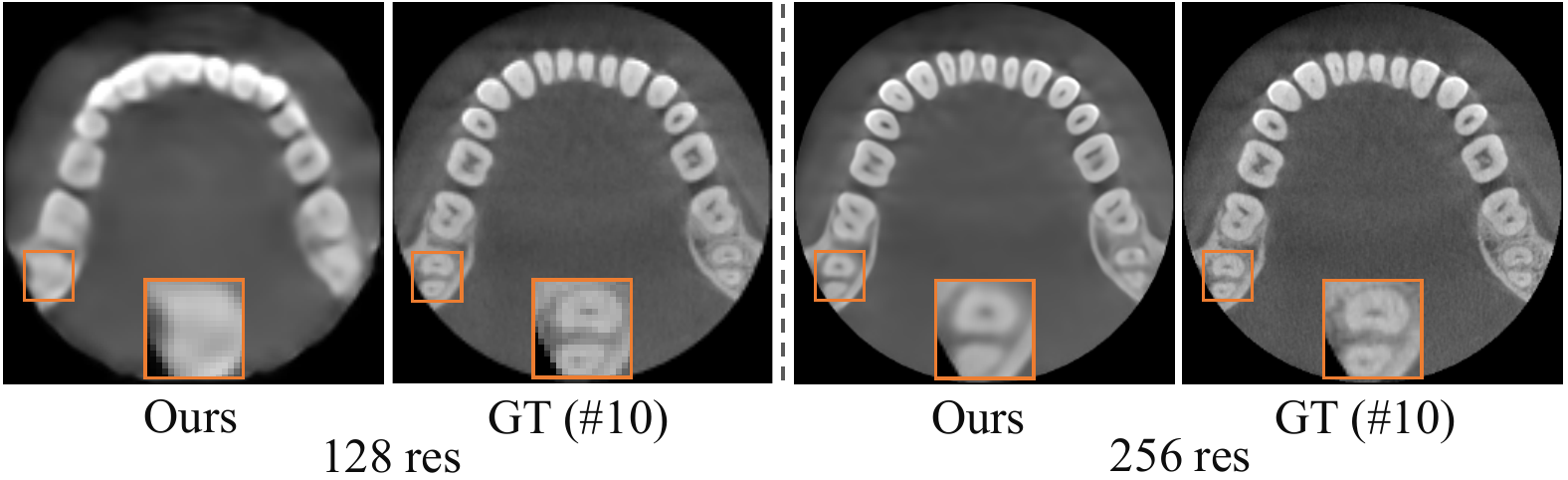}
    \caption{
        \lzt{Qualitative results of robustness analysis on reconstruction resolution on case \#10 from dental dataset (axial slice). Window: [-1000, 2000] HU.}
    }
    \label{fig: ab recon res}
\vspace{-4mm}
\end{figure}

\lzt{We first verify whether our well-trained framework can perform reconstructions at different resolutions.
We train our model with a resolution of $256^{3}$ using a downsampling rate $S=4$, and test our model with a resolution of $128^{3}$ using $S=8$.
Note that the ground truth CBCT image is also downsampled to $128^{3}$ resolution for statistical analysis.
The quantitative results are delivered in Tab.~\ref{tab: ab recon res}, and a visual example is shown in Fig.~\ref{fig: ab recon res}.
The downsampling rate $S=8$ retains only $\frac{1}{8}$ of the sample points from $S=4$ when constructing 3D feature map.
Testing with $128^{3}$ resolution leads to deteriorated results due to insufficient input information.
In conclusion, our current framework faces challenge in performing reconstructions at varying resolutions.
We would modify our training strategy by considering different resolutions to overcome it in the future.}

\subsubsection{Angle Sampling}

\begin{table}[t]
\caption{\lzt{Quantitative results of robustness analysis on angle sampling on dental dataset.}}
\centering
\setlength{\tabcolsep}{5pt}
\renewcommand\arraystretch{1.5}
\begin{tabular}{c|cc}
\hline
Test Angle                 & PSNR  & SSIM    \\ \hline
Uniform $[0^\circ, 360^\circ)$  & 31.44$\pm$1.00                & 0.891$\pm$0.016 \\
Uniform $[5^\circ, 360^\circ)$  & 30.97$\pm$0.90                & 0.884$\pm$0.016 \\
Uniform $[10^\circ, 360^\circ)$ & 30.28$\pm$0.76                & 0.873$\pm$0.016 \\
Uniform $[20^\circ, 360^\circ)$ & 29.67$\pm$0.74                & 0.863$\pm$0.016 \\
Random $[0^\circ, 360^\circ)$   & 25.04$\pm$1.76                & 0.766$\pm$0.046 \\ \hline
\end{tabular}
\label{tab: ab sampled angle}
\vspace{-2mm}
\end{table}

\begin{figure}[t]
\centering
    \includegraphics[width=0.49\textwidth]{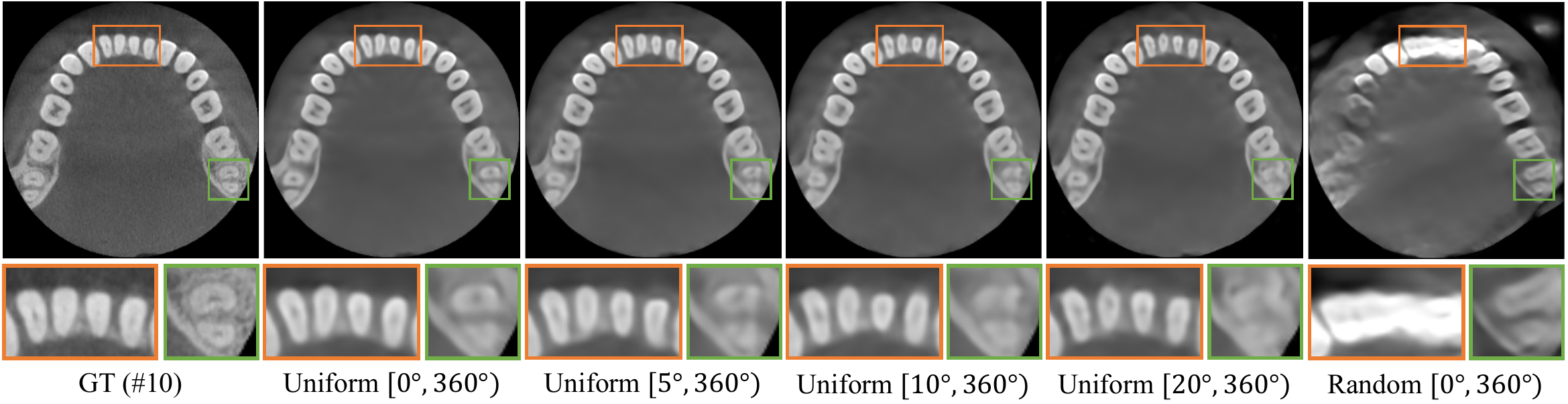}
    \caption{
        \lzt{Qualitative results of robustness analysis on angle sampling on case \#10 from dental dataset (axial slice). Window: [-1000, 2000] HU.}
    }
    \label{fig: ab sampled angle}
    \vspace{-4mm}
\end{figure}

\lzt{Second, we examine whether our framework is robust to the different viewing angles when testing.
Our model is trained with uniformly sampled angles within a range of $[0^\circ, 360^\circ)$ and a consistent angular step.
We first test with different starting angles for uniform sampling, e.g., $5^\circ$, $10^\circ$, and $20^\circ$.
Second, we test with randomly sampled angles} 
\lzt{within a range of $[0^\circ, 360^\circ)$ for inconsistent angular step.
The quantitative results are delivered in Tab.~\ref{tab: ab sampled angle}, and a visual example is shown in Fig.~\ref{fig: ab sampled angle}. 
As the starting angle increases from $0^\circ$ to $5^\circ$, $10^\circ$ and then to $20^\circ$, the reconstruction quality gradually decreases.
This decline occurs because the input view sampling angles progressively deviate from the training setting, making adaption more challenging for our model. 
When testing with random sampling, the reconstruction quality drops significantly and the metric variances increase. 
This is because the randomly sampled angles deviate considerably from the training setting.
And the angle deviation varies among different test samples, leading to increased variances.
In conclusion, our current framework is not robust to different starting angles or inconsistent angular steps, especially for random sampling angles.
We would solve it by using different angle sampling settings during training stage in the future.}

\subsubsection{Number of Input Views}

\begin{table}[th]
\caption{\lzt{Quantitative results of robustness analysis on number of input views on dental dataset.}}
\centering
\setlength{\tabcolsep}{5pt}
\renewcommand\arraystretch{1.5}
\begin{tabular}{c|cc}
\hline
Test Views & PSNR  & SSIM    \\ \hline
5 views       & 23.95$\pm$0.55                & 0.715$\pm$0.029 \\
10 views      & 28.83$\pm$1.19                & 0.850$\pm$0.025 \\
20 views      & 28.54$\pm$0.78                & 0.842$\pm$0.021 \\ \hline
\end{tabular}
\label{tab: ab number of test views}
\vspace{-2mm}
\end{table}

\begin{figure}[th]
\centering
    \includegraphics[width=0.49\textwidth]{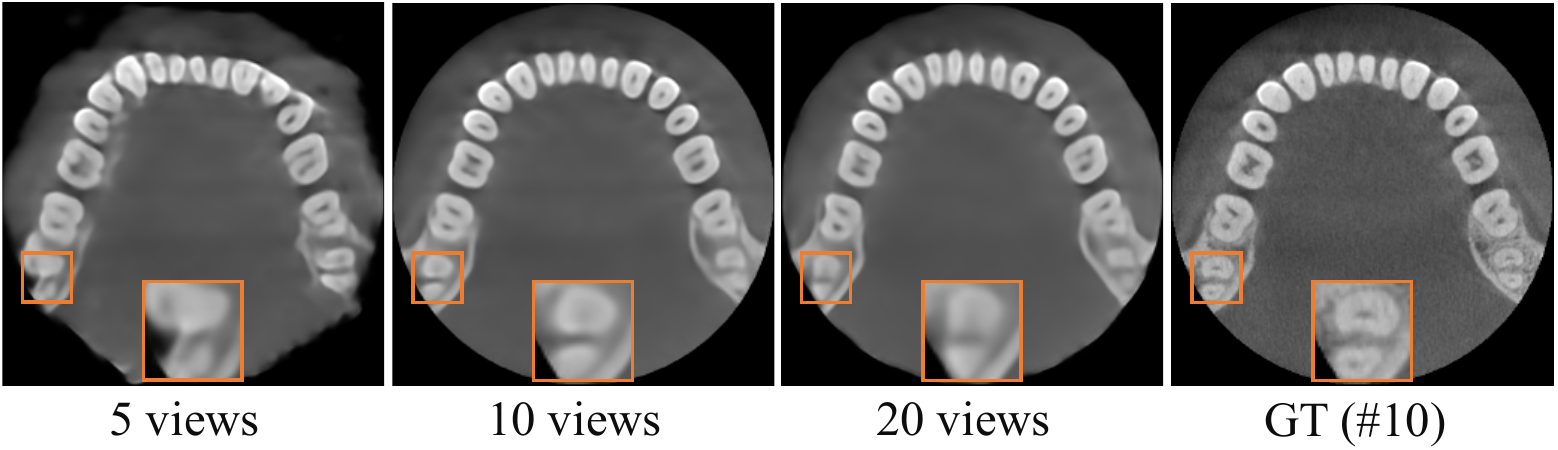}
    \caption{
        \lzt{Qualitative results of robustness analysis on number of input views on case \#10 from dental dataset (axial slice). Window: [-1000, 2000] HU.}
    }
    \label{fig: ab number of test views}
    \vspace{-4mm}
\end{figure}

\lzt{At last, we examine whether our framework is robust to the number of input views when testing.
We train our model with 10 input views, and test it with 5 or 20 views.
The quantitative results are delivered in Tab.~\ref{tab: ab number of test views}, and a visual example is shown in Fig.~\ref{fig: ab number of test views}.
According to Sec.~\ref{sec:dataacq}, the input of 20 views retains all the sampling angels from 10 views.
In contrast, the input of 5 views retains only a half of the sampling angles from 10 views.
When training with 10 views, testing with 5 views results in degradation due to insufficient sampling angles, but testing with 20 views is acceptable as 20 views contain sufficient sampling angles.
Hence, our model can still achieve satisfactory reconstruction results when the sampling angles of test input include the training ones.
Otherwise, the reconstruction quality may degrade due to insufficient input information.
Thus, our model is not robust to the number of input views.
In the future, we plan to take different number of views as input during the training stage to address it.}


\section{Discussion}

While our framework has shown promising results, it does have its limitations. 
\R{Firstly, one} significant challenge is that the method is resource-intensive and not computational memory efficient. 
\R{Secondly}, our reconstruction results for low contrast images, such as spine and walnut images, are not yet up to the mark. 
The reconstructed results often turn out too smooth, lacking finer details, 
And there can even be structural errors in walnut images with 10 or 5 input views. 
\R{Thirdly}, our current framework may be not robust to varying test conditions that differ from the training ones, including reconstruction resolution, angle sampling, and the number of input views. 
\R{At last, real-world experiments on walnut dataset do not fully reflect the real-life scenario as walnut scans lack significant scattering effects.
Throughout our paper, we do not take scattering effects into consideration and only account for the primary rays.
However, in real-life thorax and pelvic scans, scattering effects are much more severe, even producing more scatter rays than primary rays.
This usually leads to increased image artifacts and noises.
Our method's robustness in real-life clinical scenario requires further verification.}

In the future, we aim to improve our reconstruction performance by integrating explicit shape or boundary priors to help correct structural errors. 
Additionally, we plan to design robust training strategies that enable the model to adapt to varying test conditions.
\R{What's more, it would be valuable to verify our method's robustness using real-life projection data with significant scattering effects, such as scans of thorax and pelvic.
We could also incorporate scattering effects to improve our algorithm for more accurate image reconstruction in clinical practice.}


\section{Conclusion}

In this paper, we introduced a novel framework for sparse-view CBCT reconstruction. 
Our method respects the inherent nature of X-ray perspective projection during the feature \lzt{back projection}, ensuring \lzt{accurate information retrieval from multiple X-ray projections.}
Moreover, by leveraging the prior knowledge learned from our extensive dataset, our framework efficiently tackles the challenges posed by sparse-view inputs, delivering high-quality reconstructions. 
The effectiveness and \lzt{time} efficiency are thoroughly validated through extensive testing on both \lzt{simulated and real-world datasets.}

{
\bibliographystyle{ieeetr}
\normalem
\bibliography{reference}
}

\end{document}